\documentclass[aps, prd, superscriptaddress, 10pt, twocolumn]{revtex4-2}

\usepackage[english]{babel}
\usepackage[T1]{fontenc}

\usepackage{comment}
\usepackage{graphicx} 

\usepackage{amsmath}
\usepackage{amssymb}
\usepackage{mathtools}
\usepackage{mleftright} 
\usepackage{simpler-wick} 
\usepackage{slashed}

\usepackage{pgfplots}
\usepgfplotslibrary{patchplots}
\pgfplotsset{compat = 1.18}
\usepackage[compat = 1.1.0]{tikz-feynman} 
\usetikzlibrary{calc, positioning, shapes}
\usetikzlibrary{arrows.meta, backgrounds, fit, decorations.markings, decorations.pathreplacing, matrix, positioning}

\usepackage[dvipsnames]{xcolor} 

\usepackage{url}
\usepackage{hyperref}

\allowdisplaybreaks
\begin{document}

\title{Connecting baryon light-front wave functions\texorpdfstring{\\}{}to quasi-transverse-momentum-dependent correlators in lattice QCD}

\author{Simone Rodini}
\thanks{Electronic address: \href{mailto:simone.rodini@unipv.it}{simone.rodini@unipv.it}}
\affiliation{Dipartimento di Fisica "A. Volta", Università degli Studi di Pavia, I-27100 Pavia, Italy}
\affiliation{Istituto Nazionale di Fisica Nucleare, Sezione di Pavia, I-27100 Pavia, Italy}

\author{Andrea Schiavi}
\thanks{Electronic address: \href{mailto:andrea.schiavi01@universitadipavia.it}{andrea.schiavi01@universitadipavia.it}}
\affiliation{Dipartimento di Fisica "A. Volta", Università degli Studi di Pavia, I-27100 Pavia, Italy}
\affiliation{Istituto Nazionale di Fisica Nucleare, Sezione di Pavia, I-27100 Pavia, Italy}

\author{Barbara Pasquini}
\thanks{Electronic address: \href{mailto:barbara.pasquini@unipv.it}{barbara.pasquini@unipv.it}}
\affiliation{Dipartimento di Fisica "A. Volta", Università degli Studi di Pavia, I-27100 Pavia, Italy}
\affiliation{Istituto Nazionale di Fisica Nucleare, Sezione di Pavia, I-27100 Pavia, Italy}

\date{\today}

\begin{abstract}

The light-front wave functions (LFWFs) of a hadron are nonperturbative objects that encode information on the configurations of the constituent partons. We show how to extract the LFWFs of baryons, such as the proton, from equal-time correlators suitable for Lattice QCD simulations. Using an operator product expansion, we prove the factorization of the relevant correlator in the three-quark color-singlet LFWF, a residual lattice factor, and a soft factor that systematically subtracts the additional divergences arising from the factorization. We verify up to next-to-leading order the independent renormalizability of the LFWF, and we derive the evolution equations that govern its scale dependence.

\end{abstract}
\vspace*{-3\baselineskip}

\maketitle

\section{Introduction}
\label{intro_sec}

Quantum Chromodynamics (QCD) provides the fundamental description of the strong interaction between quarks and gluons, collectively referred to as partons. Due to the confinement of color charges, these degrees of freedom cannot be observed as free particles, but only as constituents of color-singlet bound states called hadrons. While asymptotic freedom allows for a perturbative treatment of QCD at high energies, it remains an open problem the nonperturbative dynamics, which is responsible for the formation and internal structure of the hadrons.

Light-front wave functions (LFWFs) are quantum amplitudes that connect a hadron state to the vacuum through an operator defined on a hypersurface tangent to the light cone, which represents a partonic configuration in the hadron. The LFWFs encode the full momentum-space structure of the partonic configurations, and, in principle, contain all the information on the internal dynamics of the bound states. In particular, experimentally measurable quantities such as parton distribution functions can be described as products of LFWFs where some partonic degrees of freedom are integrated out. These distributions therefore focus on specific aspects of the more complete information contained in the LFWFs, which, in this sense, represent the most fundamental nonperturbative quantities describing hadron structure.

In this paper, we work directly on the LFWFs as the central nonperturbative objects. To access them, we turn to Lattice QCD, which allows for first-principle nonperturbative calculations of hadronic matrix elements by formulating QCD in Euclidean spacetime on a bounded and discrete grid. Lattice simulations give access to equal-time correlators, therefore dedicated factorization theorems are required to connect them to the light-front amplitudes we are interested in. The factorization produces additional  divergences that must cancel in physical observables, and requires the nonperturbative quantities to be renormalizable independently of lattice artefacts. Factorization frameworks have been developed for transverse-momentum-dependent (TMD) parton distribution and fragmentation functions~\cite{Vladimirov:2020ofp, Rodini:2022wic} as well as generalized parton distributions (see, e.g., reviews~\cite{Cichy:2018mum, Ji:2020ect, Constantinou:2020pek, Lin:2025hka} and references therein), and for baryon distribution amplitudes~\cite{Deng:2023csv} and meson LFWFs~\cite{Ji:2021znw, Deng:2022gzi, LatticeParton:2023xdl}. In this work, we extend this program to the LFWFs of baryons, most notably the proton, isolating the relevant partonic degrees of freedom through a TMD operator expansion based on the background field method~\cite{Abbott:1981ke, Vladimirov:2021hdn}.

The paper is organized as follows. In Sec.~\ref{qtmd_correlator_sec}, we define the LFWF of a baryon and the corresponding equal-time correlator on the lattice, called a quasi-transverse-momentum-dependent (QTMD) correlator, and present the TMD expansion at leading power. In Sec.~\ref{factor_qtmd_correlator_sec}, we analyze the structure of divergences of the factorized QTMD correlator, which is composed of the three-quark baryon LFWF, a residual lattice factor, and a soft factor to compensate the additional divergences arising from the factorization~\cite{Echevarria:2015byo, Vladimirov:2017ksc}. Alongside ultraviolet divergences, there are rapidity divergences coming from the interaction of the original fields with lightlike gauge links at infinity. In Sec.~\ref{cancel_divs_n_physical_lfwf_sec}, we prove the cancellation of these divergences at next-to-leading order, and the independent renormalizability of the physical LFWF. This is a function of one ultraviolet renormalization scale, alongside one rapidity scale for each quark, with independent evolution with respect to each of them, which we verify in Sec.~\ref{scale_evo_sec}. In Sec.~\ref{conclusions_sec}, we summarize our results, and conclude by commenting on possible future developments.

\section{The QTMD Correlator}
\label{qtmd_correlator_sec}

\begin{figure}[ht]
\centering
\includegraphics[scale = 0.83]{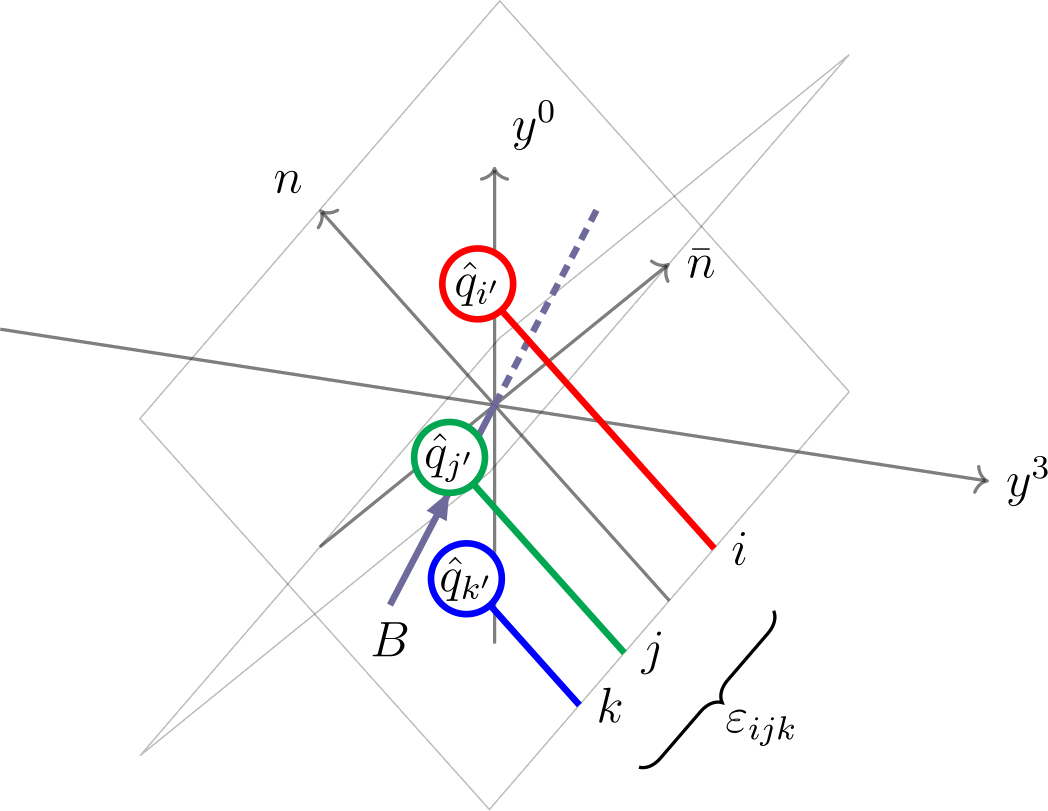}
\hspace*{1cm}
\includegraphics[scale = 0.83]{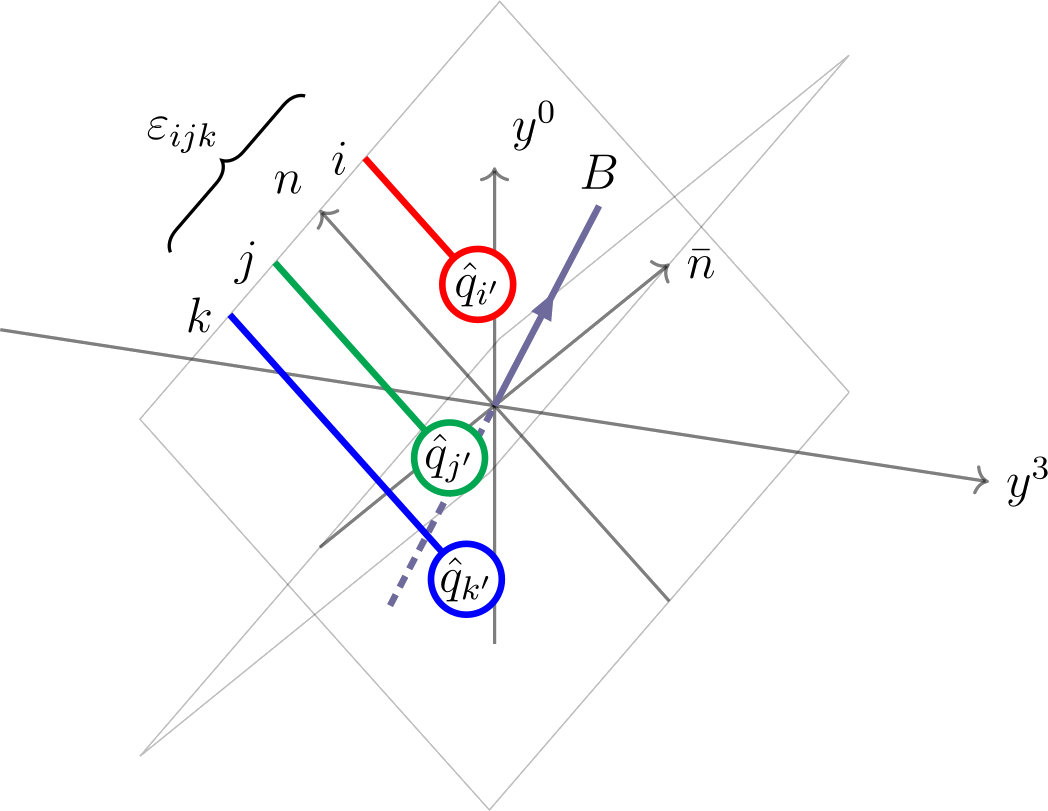}
\caption{Three-quark baryon LFWF. The quark fields lie on the hypersurface tangent to the light cone and orthogonal to $ \bar{ n } $. The lightlike Wilson lines to infinity are colored in red, green and blue to visually indicate the antisymmetrization of the fundamental-representation color indices. For simplicity, the transverse gauge links are implied. The figure on the left corresponds to a baryon in the initial state, and the figure on the right to a baryon in the final state.}
\label{inoutbaryon_3q_lfwf_fig}
\end{figure}

We define the three-quark color-singlet LFWF of a baryon $ B $ with four-momentum $ P $ and spin $ S $ as
\begin{widetext}
\begin{alignat}{1}
  \widetilde{ \Phi }_{ 3 q; B } \! \left( y_{ 1 }, y_{ 2 }, y_{ 3 } \right) = \langle 0 \rvert \varepsilon_{ i j k }
& \left[ \pm \infty n + \infty_{ \perp }, \pm \infty n + b_{ 1 } \right]_{ i i'' } \left[ \pm \infty n + b_{ 1 }, y_{ 1 }^{ - } n + b_{ 1 } \right]_{ i'' i' } \nonumber \\
  \times
& \left[ \pm \infty n + \infty_{ \perp }, \pm \infty n + b_{ 2 } \right]_{ j j'' } \left[ \pm \infty n + b_{ 2 }, y_{ 2 }^{ - } n + b_{ 2 } \right]_{ j'' j' } \nonumber \\
  \times
& \left[ \pm \infty n + \infty_{ \perp }, \pm \infty n + b_{ 3 } \right]_{ k k'' } \left[ \pm \infty n + b_{ 3 }, y_{ 3 }^{ - } n + b_{ 3 } \right]_{ k'' k' } \nonumber \\
  \times
& \hat{ q }_{ i' } \! \left( y_{ 1 }^{ - } n + b_{ 1 } \right) \hat{ q }_{ j' } \! \left( y_{ 2 }^{ - } n + b_{ 2 } \right) \hat{ q }_{ k' } \! \left( y_{ 3 }^{ - } n + b_{ 3 } \right) \lvert B \! \left( P, S \right) \rangle ,
\label{baryon_lfwf}
\end{alignat}
\end{widetext}
where, using decomposition (\ref{sudakov_decomposition}) of a four-vector in two lightlike directions $ \bar{ n }, n $ and a remaining transverse vector, for every $ q = 1, 2, 3 $ we have $ n y_{ q } = y_{ q }^{ + } = 0 $. The $ \hat{ q } $ are quark field operators, and from now on we will suppress the hat-notation for operators in formulas involving their matrix elements. The quark fields are antisymmetrized over their fundamental-representation color indices through Wilson lines defined as
\begin{equation}
\left[ y_{ \text{fin} }, y_{ \text{in} } \right] = P \! \left[ \exp \! \left( i g \int_{ y_{ \text{in} } }^{ y_{ \text{fin} } } \! dy^{ \mu } A_{ \mu } \! \left( y \right) \right) \right] \! ,
\label{wilson_line_def}
\end{equation}
with $ g $ the strong coupling constant, $ A_{ \mu } $ the gluon field operator, and $ P \! \left[ ... \right] $ the path-ordering operator. The $ n $-directed Wilson lines go to $ - \infty, + \infty $ for a baryon in the initial or final state, respectively, with the two cases represented in Fig. \ref{inoutbaryon_3q_lfwf_fig}.

\begin{figure*}[ht]
\centering
\includegraphics[scale = 0.83]{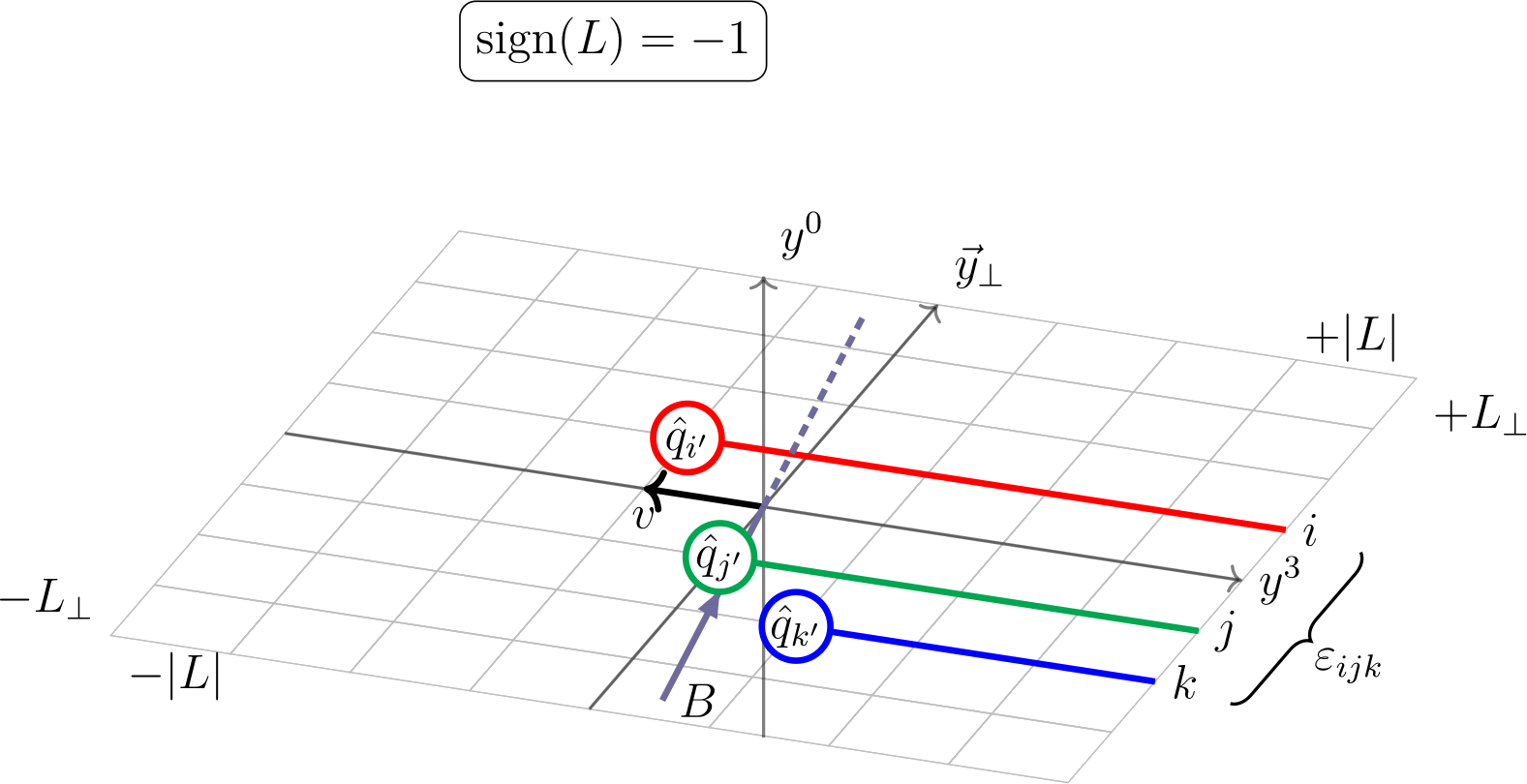}\\[1cm]
\includegraphics[scale = 0.83]{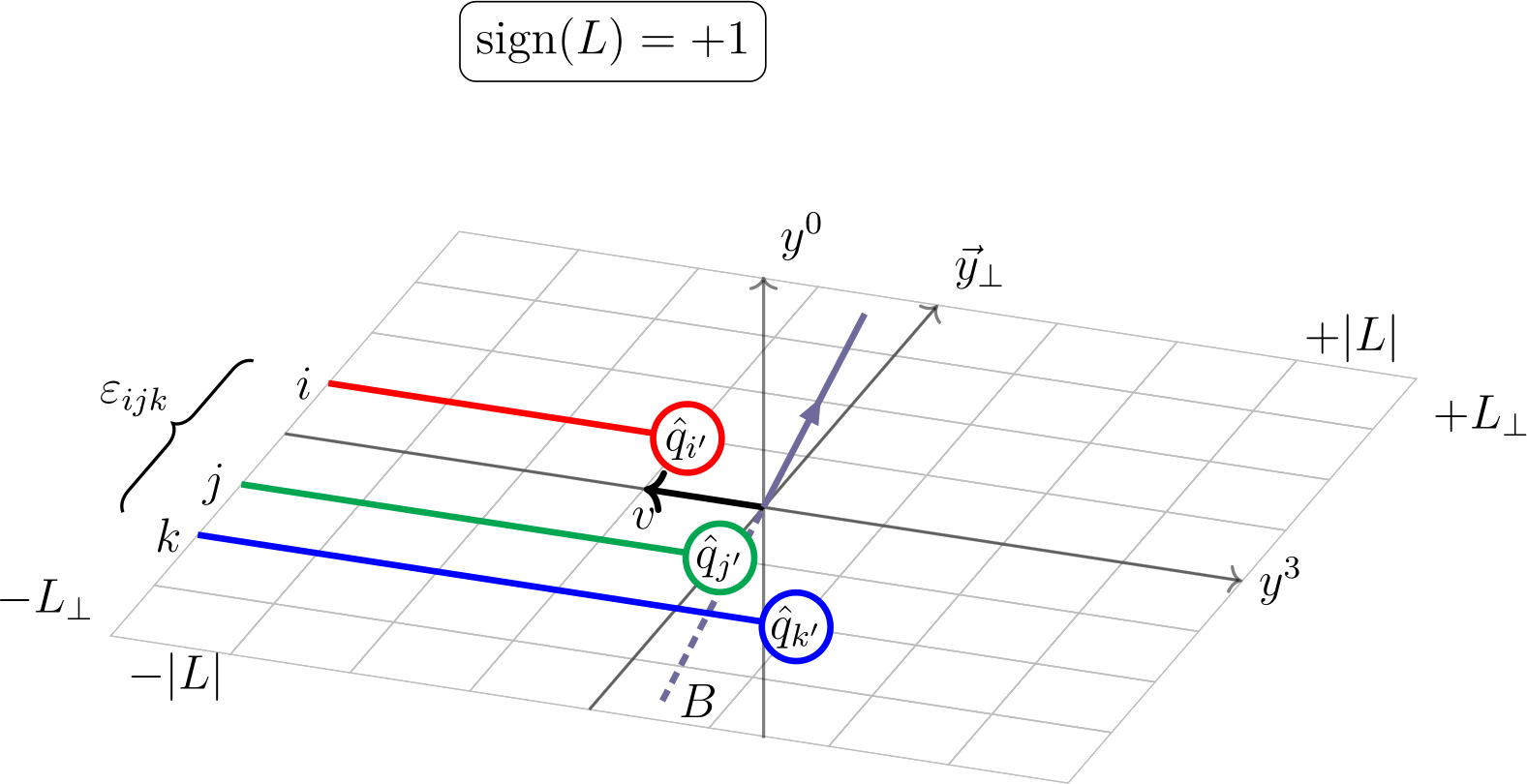}
\caption{Baryon QTMD correlator. We consider the baryon $ B $ traveling along the $ y^{ 3 } $-axis, and the quark operators $ \hat{ q } $ on the hypersurface $ y^{ 0 } = 0 $ are connected to the boundary of the lattice through Wilson lines. The lattice is bounded by $ \lvert L \rvert $ along the $ y^{ 3 } $-axis, and by $ L_{ \perp } $ along the perpendicular spatial directions. The cases $ \text{sign} \! \left( L \right) = - 1, + 1 $ correspond to a baryon in the initial and final state, respectively. For simplicity, we only represent the $ v $-directed gauge links, colored in red, green and blue to visually indicate the antisymmetrization of the color indices in the fundamental representation.}
\label{inoutbaryon_qtmd_correlator_fig}
\end{figure*}

To allow for simulations within the framework of Lattice QCD, alongside the baryon LFWF we define a QTMD correlator as
\begin{equation}
\widetilde{ \Omega }_{ v; B } \! \left( \left\{ y_{ q } \right\} \right) = \langle 0 \rvert \varepsilon_{ i j k } J_{ v, i } \! \left( y_{ 1 } \right) J_{ v, j } \! \left( y_{ 2 } \right) J_{ v, k } \! \left( y_{ 3 } \right) \! \lvert B \! \left( P, S \right) \rangle ,
\label{baryon_qtmd_correlator_def}
\end{equation}
where now
\begin{equation}
y_{ q } = l_{ q } v + b_{ q },
\label{quasi_initial_position}
\end{equation}
and, without loss of generality, we can choose $v$ as
\begin{equation}
v = \frac{ n - \bar{ n } }{ \sqrt{ 2 } } = \frac{ 1 }{ \sqrt{ 2 } } \left( - 1, 1, 0, 0 \right) \! .
\label{v_direction}
\end{equation}
The current $ J $ is defined as
\begin{equation}
J_{ v } \! \left( y \right) = \left[ Lv + L_{ \perp }, Lv + y_{ \perp } \right] \left[ Lv + y_{ \perp }, y \right] q \! \left( y \right) \! ,
\label{quark_op_vwilsoned_to_infinity}
\end{equation}
where $ L, L_{ \perp } $ are finite on the lattice, but are assumed to be much bigger than every $ l , b $, allowing us to treat them effectively as infinite in calculations. For this reason, the current is referred to as a semicompact operator. We consider $ \text{sign} \! \left( L \right) = - 1, + 1 $ for a hadron in the initial or final state, respectively, with the two versions of the baryon QTMD correlator represented in Fig. \ref{inoutbaryon_qtmd_correlator_fig}.
Since the QTMD correlator is defined at equal time, it can be directly formulated on a Euclidean lattice; moreover, it can be expressed as a path integral, i.e.,
\begin{alignat}{1}
  \widetilde{ \Omega }_{ v; B } \! \left( y_{ 1 }, y_{ 2 }, y_{ 3 } \right) = \int
& \! \left[ Dq D \! A \right] e^{ i S_{ \text{QCD} } \left[ q, A \right] } \phi \! \left( P, S \right) \nonumber \\
  \times
& \varepsilon_{ i j k } J_{ v, i } \! \left( y_{ 1 } \right) J_{ v, j } \! \left( y_{ 2 } \right) J_{ v, k } \! \left( y_{ 3 } \right) \! ,
\label{baryon_qtmd_correlator_path_integral}
\end{alignat}
where $ S_{ \text{QCD} } $ is the QCD action, and $ \phi $ is the baryon field. The latter is not known, but its explicit expression will not be required in what follows. In the rest of the paper, we will show at next-to-leading order the factorization of the three-quark LFWF from the QTMD correlator in the limit of large baryon momentum.

Consider
\begin{equation}
P = P^{ + } \bar{ n } + \frac{ M^{ 2 } }{ 2 P^{ + } } n,
\label{high_collinear_momentum}
\end{equation}
where $ M \ll P^{ + } $ is the mass of the baryon. Using the background field method~\cite{Abbott:1981ke} with two background fields adapted to a TMD expansion~\cite{Vladimirov:2021hdn}, we shift the fields as
\begin{equation}
q \mapsto \psi + q_{ \bar{ n } } + q_{ v }, \qquad A^{ \mu } \mapsto B^{ \mu } + A^{ \mu }_{ \bar{ n } } + A^{ \mu }_{ v },
\label{background_shifted_fields}
\end{equation}
where $ \psi, B^{ \mu } $ are the dynamical fields, while the remaining ones are treated as background. The $ \bar{ n } $-collinear fields, which are almost collinear to the baryon, are required to satisfy the scaling relations
\begin{equation}
\left( \partial^{ + }, \partial^{ - }, \partial_{ \perp } \right) \! \left( q_{ \bar{ n } }, A_{ \bar{ n } } \right) \lesssim \left( 1, \lambda^{ 2 }, \lambda \right) \! P^{ + } \! \left( q_{ \bar{ n } }, A_{ \bar{ n } } \right) \! ,
\label{nbar_field_derivatives_scaling}
\end{equation}
with
\begin{equation}
\lambda = \left( M / P^{ + } \right) \ll 1.
\label{smallness_param}
\end{equation}
Central to our study is the ability to retain the dependence on the transverse momentum of the partons. To this end, we allow the transverse field separations inside the baryon to be large, namely of the order of $ b \sim ( 1 / M ) = ( \lambda P^{ + } )^{ -1 } $. Therefore, terms of the form $ b^{ \mu } \partial_{ \mu } \! \left( q_{ \bar{ n } }, A_{ \bar{ n } } \right) \sim \left( q_{ \bar{ n } }, A_{ \bar{ n } } \right) $ cannot be ignored. We still require a suppression of the anti-collinear momentum, i.e., $ y^{ + } \partial^{ - } \! \left( q_{ \bar{ n } }, A_{ \bar{ n } } \right) \sim \lambda \! \left( q_{ \bar{ n } }, A_{ \bar{ n } } \right) $ or smaller. At the same time, to stay away from the small-$ x $ regime, where $ x $ is the fraction of longitudinal momentum of the baryon carried by a parton, we want to keep $ y^{ - } \partial^{ + } \! \left( q_{ \bar{ n } }, A_{ \bar{ n } } \right) \sim \left( q_{ \bar{ n } }, A_{ \bar{ n } } \right) $. Since $ y^{ + } \sim l \sim y^{ - } $, it must be
\begin{equation}
\left( l, b \right) \sim \frac{ 1 }{ P^{ + } } \! \left( 1, \frac{ 1 }{ \lambda } \right) \! .
\label{y_scaling}
\end{equation}
Using the projectors defined in Eqs. (\ref{lc_good_projector}), (\ref{lc_bad_projector}), we define the light-cone (LC) good and bad components of the collinear quark field as, respectively,
\begin{alignat}{1}
  \xi_{ \bar{ n } } 
& = \Lambda_{ + } q_{ \bar{ n } }, 
\label{collinear_lc_good_component} \\
  \eta_{ \bar{ n } } 
& = \Lambda_{ - } q_{ \bar{ n } }.
\label{collinear_lc_bad_component}
\end{alignat}
From the Dirac equation, we have
\begin{equation}
\eta_{ \bar{ n } } \sim \lambda \xi_{ \bar{ n } },
\label{collinear_lc_good_vs_bad_scaling}
\end{equation}
while the gluon field is a vector with natural dimensions of momentum, therefore it scales as its momentum. The $ \bar{ n } $-collinear fields do not encode all the nonperturbative physics, hence we also need the $ v $-collinear fields, defined by the scaling relations
\begin{equation}
\left( \partial^{ + }, \partial^{ - }, \partial_{ \perp } \right) \! \left( q_{ v }, A_{ v } \right) \lesssim \left( \lambda^{ 2 }, \lambda^{ 2 }, \lambda \right) \! P^{ + } \! \left( q_{ v }, A_{ v } \right) \! .
\label{v_field_derivatives_scaling}
\end{equation}
The region of small longitudinal momentum is accounted for by both the $ \bar{ n } $-collinear and $ v $-collinear fields. Outside the small-$ x $ regime, we can assume that hadrons do not contain soft partons, therefore the overlap region reduces to a vacuum contribution, known as the soft factor $ S \! \left( \left\{ y \right\} \right) $. Therefore, to remove the double-counting of the overlap region, we can define
\begin{widetext}
\begin{equation}
\widetilde{ \Omega }_{ v; B } \! \left( y_{ 1 }, y_{ 2 }, y_{ 3 } \right) = \int \! \left[ D q_{ \bar{ n } } D \! A_{ \bar{ n } } \right] \! \left[ D q_{ v } D \! A_{ v } \right] e^{ i S_{ \text{QCD} } \left[ q_{ \bar{ n } }, A_{ \bar{ n } } \right] } e^{ i S_{ \text{QCD} } \left[ q_{ v }, A_{ v } \right] } \frac{ O_{ \text{eff} } \! \left( y_{ 1 }, y_{ 2 }, y_{ 3 } \right) }{ S \! \left( y_{ 1 }, y_{ 2 }, y_{ 3 } \right) } \phi \! \left( P, S \right) \! ,
\label{baryon_qtmd_correlator_path_integral_with_bkg}
\end{equation}
where
\begin{equation}
O_{ \text{eff} } \! \left( y_{ 1 }, y_{ 2 }, y_{ 3 } \right) = \int \! \left[ D \psi D \! B \right] e^{ i S_{ \text{int} } \left[ \psi, q_{ \bar{ n } }, q_{ v }, B, A_{ \bar{ n } }, A_{ v } \right] } \varepsilon_{ i j k } J_{ v, i } \! \left( y_{ 1 } \right) J_{ v, j } \! \left( y_{ 2 } \right) J_{ v, k } \! \left( y_{ 3 } \right) \! ,
\label{effective_baryon_qtmd_operator_with_bkg}
\end{equation}
with
\begin{equation}
S_{ \text{int} } \! \left[ \psi, q_{ \bar{ n } }, q_{ v }, B, A_{ \bar{ n } }, A_{ v } \right] = S_{ \text{QCD} } \! \left[ \psi + q_{ \bar{ n } } + q_{ v }, B + A_{ \bar{ n } } + A_{ v } \right] - S_{ \text{QCD} } \! \left[ q_{ \bar{ n } }, A_{ \bar{ n } } \right] - S_{ \text{QCD} } \! \left[ q_{ v }, A_{ v } \right] \! .
\label{interaction_action_with_bkg}
\end{equation}
\end{widetext}
We postpone to Sec.~\ref{factor_qtmd_correlator_sec} the definition of the appropriate soft factor, which, as we will see in Sec.~\ref{cancel_divs_n_physical_lfwf_sec}, will compensate the divergences arising from the separation of the $ \bar{ n } $-collinear and $ v $-collinear sectors, allowing us to extract the physical three-quark baryon LFWF from the QTMD correlator.

Using the scaling relation (\ref{y_scaling}), interactions between volumes around the positions of the quark fields are mediated by propagators of order $ b^{ - 2 } \sim \left( \lambda P^{ + } \right)^{ 2 } $. These contributions are not relevant for leading-power (LP) and next-to-leading-power (NLP) calculations. Therefore, the LP and NLP functional integral in Eq. (\ref{effective_baryon_qtmd_operator_with_bkg}) factorizes around the positions of the quarks, i.e.,
\begin{equation}
O_{ \text{eff} } \! \left( y_{ 1 }, y_{ 2 }, y_{ 3 } \right) = \varepsilon_{ i j k } \mathcal{ J }_{ v, i } \! \left( y_{ 1 } \right) \mathcal{ J }_{ v, j } \! \left( y_{ 2 } \right) \mathcal{ J }_{ v, k } \! \left( y_{ 3 } \right) \! ,
\label{upto_nlp_effective_baryon_qtmd_operator_with_bkg}
\end{equation}
where we introduced the effective current
\begin{equation}
\mathcal{ J } \! \left( y \right) = \int \! \left[ D \psi D \! B \right] e^{ i S_{ \text{int} } \left[ \psi, q_{ \bar{ n } }, q_{ v }, B, A_{ \bar{ n } }, A_{ v } \right] } J_{ v } \! \left( y \right) \! ,
\label{effective_quark_current}
\end{equation}
with integrals restricted around $ y $.

After integrating over the dynamical fields, the leading-order (LO) contribution is given by tree-level diagrams. To further develop the effective current, we can leverage the freedom in fixing a gauge. The choice for the background fields can be made independently of that for the dynamical ones if we use the background-Feynman gauge for the latter (see Ref.~\cite{Abbott:1981ke}). For the former we choose the LC gauge
\begin{equation}
A_{ \bar{ n } }^{ + } = 0, \qquad A_{ v }^{ - } = 0,
\label{lc_gauge_for_bkg_fields}
\end{equation}
which does not completely fix the gauge, and we use the residual freedom to also set to zero the transverse components at infinity, specifically
\begin{equation}
\lim_{ z^{ - } \to \text{sign} \left( L \right) \infty } A_{ \bar{ n }, \perp }^{ \mu } \! \left( z \right) = 0, \qquad \lim_{ z^{ + } \to \text{sign} \left( L \right) \infty } A_{ v, \perp }^{ \mu } \! \left( z \right) = 0.
\label{residual_lc_gauge_for_bkg_fields}
\end{equation}
These choices reduce the transverse Wilson lines to the identity, and the remaining gauge links are the identity up to NLP corrections. However, in order to extract the three-quark baryon LFWF in the end, we must keep the representation of the $ v $-collinear sector in terms of the Wilson line
\begin{equation}
P \! \left[ \exp \! \left( i g \int_{ 0 }^{ L } \! d \sigma v_{ \mu } A_{ v }^{ \mu } \! \left( \sigma v + y \right) \right) \right] = H^{ \dagger } \! \left( y \right) = H^{ \dagger } \! \left( y_{ \perp } \right) \! .
\label{vcollinear_vwilson_line_def}
\end{equation}
The last equality holds up to next-to-leading power, after expanding the fields and using the scaling relation (\ref{v_field_derivatives_scaling}). Therefore, the LP and LO contribution to the effective current (\ref{effective_quark_current}) is
\begin{equation}
\mathcal{ J } \! \left( y \right) = H^{ \dagger } \! \left( y_{ \perp } \right) \xi_{ \bar{ n } } \! \left( y \right) \! .
\label{lp_lo_effective_quark_current}
\end{equation}

\begin{figure}[ht]
\centering
\includegraphics[scale = 1]{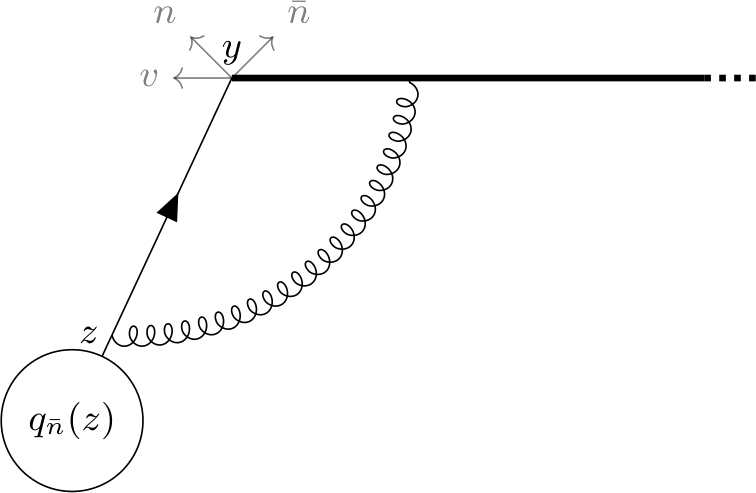}
\caption{One-loop diagram contributing to the quark current at next-to-leading order. The thick line is the gauge link.}
\label{lp_nlo_effective_quark_current_fig}
\end{figure}

In general, we have
\begin{equation}
\mathcal{ J } \! \left( y \right) = H^{ \dagger } \! \left( y_{ \perp } \right) \widehat{ C }_{ 1 } \xi_{ \bar{ n } } \! \left( y \right), \qquad \widehat{ C }_{ 1 } = I + \mathcal{ O } \! \left( \alpha_{ s } \right) \! ,
\label{lp_effective_quark_current_corrected}
\end{equation}
where the coefficient functions (CFs) are integral operators acting only on the collinear fields, and $ I $ is the identity. The next-to-leading-order (NLO) correction is the contribution of a loop between the quark field and its Wilson line, see Fig. \ref{lp_nlo_effective_quark_current_fig}.

To calculate the NLO effective current (\ref{effective_quark_current}), we need the explicit expression of the interaction action (\ref{interaction_action_with_bkg}) with two background fields, which can be found in Appendix A of Ref.~\cite{Vladimirov:2021hdn}. We expand a dynamical Wilson line in the current (\ref{quark_op_vwilsoned_to_infinity}) to first order in $ g $, we pick up a $ \overline{ \psi } B q_{ \bar{ n } } $-vertex from $ S_{ \text{int} } $, and contract the corresponding fields. In $ D = 4 - 2 \epsilon $ dimensions, this gives
\begin{equation}
\wick{ i g \int_{ 0 }^{ L } \! d \sigma  v^{ \mu } \c3{ B }_{ \mu } \! \left( \sigma v + b \right) \c2{ \psi } \! \left( y \right) i g \int \! d^{ D } z \c2{ \overline{ \psi } } \c3{ \slashed{ B } } q_{ \bar{ n } } \! \left( z \right) \! . }
\label{lp_nlo_effective_quark_current_contraction}
\end{equation}
Projecting with $ \Lambda_{ + } $ as in Eq. (\ref{lc_good_projector}) to retain only LP contributions, and using Eqs. (\ref{dirac_propagator_position_space})--(\ref{as}), we have
\begin{alignat}{1}
& \widehat{ C }_{ 1, \text{NLO} } \xi_{ \bar{ n } } \! \left( y \right) = \frac{ i g^{2} C_{ F } }{ 8 \pi^{ D } } \Gamma \! \left( 1 - \epsilon \right) \Gamma \! \left( 2 - \epsilon \right) \int_{ 0 }^{ L } \! d \sigma \int \! d^{ D } z \nonumber \\
& \times \frac{ v^{ \mu } \Lambda_{ + } \left( \slashed{ y } - \slashed{ z } \right) \gamma_{ \mu } \xi_{ \bar{ n } } \! \left( z \right) }{ \left( - \left( \sigma v + y - z \right)^{ 2 } + i 0 \right)^{ 1 - \epsilon } \left( - \left( y - z \right)^{ 2 } + i 0 \right)^{ 2 - \epsilon } }.
\label{lp_nlo_cf}
\end{alignat}
After integrating, we find (see also Ref.~\cite{Rodini:2022wic})
\begin{alignat}{1}
  \widehat{ C }_{ 1, \text{NLO} } \xi_{ \bar{ n } } \! \left( y \right) ={}
& 2 a_{ s } \left( \frac{ - v^{ 2 } }{ 4 } \right)^{ \! \epsilon } C_{ F } \frac{ 1 - \epsilon }{ 1 - 2 \epsilon } \Gamma \! \left( - \epsilon \right)\nonumber \\
& \times \int_{ 0 }^{ L } \! d \tau \frac{ \left( \tau^{ 2 } \right)^{ \epsilon } }{ \tau } \xi_{ \bar{ n } } \! \left( \tau v^{ - } + y^{ - } + y_{ \perp } \right) \! .
\label{lp_nlo_cf_final}
\end{alignat}

To move to longitudinal momentum space, we Fourier transform $ l v^{ - } $ to $ x P^{ + } $. Since $ x $ is the fraction of longitudinal momentum of the baryon carried by an extracted parton, we have $ 0 \leq x \leq 1 $. However, we do the calculations more generally, allowing $ x $ to also take negative values, which corresponds to the absorption of an antiparton. Therefore, we have
\begin{alignat}{2}
&
&
& v^{ - } \int \! d l e^{ i l v^{ - } x P^{ + } } 2 a_{ s } \left( \frac{ - v^{ 2 } }{ 4 } \right)^{ \! \epsilon } C_{ F } \frac{ 1 - \epsilon }{ 1 - 2 \epsilon } \Gamma \! \left( - \epsilon \right) \nonumber \\
&
&
& \times \int_{ 0 }^{ L } \! d \tau \frac{ \left( \tau^{ 2 } \right)^{ \epsilon } }{ \tau } \xi_{ \bar{ n } } \! \left( \left( \tau + l \right) v^{ - } + b \right) \nonumber \\
  ={}
&
&
& 2 a_{ s } \left( \frac{ - v^{ 2 } }{ 4 } \right)^{ \! \epsilon } C_{ F } \frac{ 1 - \epsilon }{ 1 - 2 \epsilon } \Gamma \! \left( - \epsilon \right) \xi_{ \bar{ n } } \! \left( x, b \right) \nonumber \\
&
&
& \times \int_{ 0 }^{ L } \! \frac{ d \tau }{ \tau } \left( \tau^{ 2 } \right)^{ \epsilon } e^{ - i \tau v^{ - } x P^{ + } } \nonumber \\
  ={}
&
&
& 2 a_{ s } C_{ F } \frac{ 1 - \epsilon }{ 1 - 2 \epsilon } \Gamma \! \left( - \epsilon \right) \Gamma \! \left( 2 \epsilon \right) \xi_{ \bar{ n } } \! \left( x, b \right) \nonumber \\
&
&
& \times \left( \frac{ - v^{ 2 } }{ \left( i s s_{ x } 2 \lvert x \rvert v^{ - } P^{ + } \right)^{ 2 } } \right)^{ \! \epsilon } \nonumber \\
  \coloneq{}
&
&
& C_{ 1, \text{NLO} } \! \left( x \right) \xi_{ \bar{ n } } \! \left( x, b \right) \! .
\label{lp_nlo_cf_longitudinal_momentum_space}
\end{alignat}
In the first equality, we defined
\begin{equation}
\xi_{ \bar{ n } } \! \left( x, b \right) = v^{ - } \int \! d l e^{ i l v^{ - } x P^{ + } } \xi_{ \bar{ n } } \! \left( l v^{ - } + b \right) \! ,
\label{longitudinal_ft_of_quark_field}
\end{equation}
while in the second equality we defined $ s = \text{sign} \! \left( L \right) $, $ s_{ x } = \text{sign} \! \left( x \right) $. Technically, in Eq. (\ref{longitudinal_ft_of_quark_field}) we still have to respect $ \lvert l \rvert \ll \lvert L \rvert $, which implies $ \lvert x \rvert \gg \lvert L P^{ + } \rvert^{ - 1 } $. This condition is compatible with the assumption of being outside the small-$ x $ regime.

\section{Factorization of the QTMD Correlator}
\label{factor_qtmd_correlator_sec}

We can now show how the three-quark baryon LFWFs emerge from the QTMD correlator \eqref{baryon_qtmd_correlator_def} at leading power. Inserting back the effective quark currents (\ref{lp_effective_quark_current_corrected}) into Eq. (\ref{baryon_qtmd_correlator_path_integral_with_bkg}), and assuming that the hadron is made up of collinear partons only, the path integrals for the $ \bar{ n } $-collinear and the $ v $-collinear sectors factorize from each other. In a gauge-invariant formulation, and requiring the matrix elements to remain color neutral, we have
\begin{align}
& \widetilde{ \Omega }_{ v; B } \! \left( y_{ 1 }, y_{ 2 }, y_{ 3 } \right) \nonumber \\
& = \frac{ \Psi \! \left( b_{ 1 }, b_{ 2 }, b_{ 3 } \right) \widehat{ C }_{ 1 } \widehat{ C }_{ 1 } \widehat{ C }_{ 1 } \widetilde{ \Phi }_{ 1 1 1 } \! \left( y_{ 1 }, y_{ 2 }, y_{ 3 } \right) }{ S \! \left( b_{ 1 }, b_{ 2 }, b_{ 3 } \right) } ,
\label{lp_baryon_qtmd_correlator_factorized}
\end{align}
where $ \widetilde{ \Phi }_{ 1 1 1 } $ is the three-quark baryon LFWF (\ref{baryon_lfwf}) at leading power, $ \Psi $ is a residual lattice factor, and $ S $ is the soft factor.

\begin{widetext}
For the LFWF, explicitly we have
\begin{alignat}{3}
  \widetilde{ \Phi }_{ 1 1 1 } \! \left( y_{ 1 }, y_{ 2 }, y_{ 3 } \right) = \langle 0 \rvert \varepsilon_{ i j k }
& \left[ L n + \infty_{ \perp }, L n + b_{ 1 } \right]_{ \bar{ n }, i i' }
&
& \left[ L n + b_{ 1 }, y_{ 1 }^{ - } n + b_{ 1 } \right]_{ \bar{ n }, i' i'' }
&
& \xi_{ \bar{ n }, i'' } \! \left( y_{ 1 }^{ - } n + b_{ 1 } \right) \nonumber \\
  \times
& \left[ L n + \infty_{ \perp }, L n + b_{ 2 } \right]_{ \bar{ n }, j j'}
&
& \left[ L n + b_{ 2 }, y_{ 2 }^{ - } n + b_{ 2 } \right]_{ \bar{ n }, j' j'' }
&
& \xi_{ \bar{ n }, j'' } \! \left( y_{ 2 }^{ - } n + b_{ 2 } \right) \nonumber \\
  \times
& \left[ L n + \infty_{ \perp }, L n + b_{ 3 } \right]_{ \bar{ n }, k k'}
&
& \left[ L n + b_{ 3 }, y_{ 3 }^{ - } n + b_{ 3 } \right]_{ \bar{ n }, k' k'' }
&
& \xi_{ \bar{ n }, k'' } \! \left( y_{ 3 }^{ - } n + b_{ 3 } \right) \! \lvert B \! \left( P, S \right) \rangle ,
\label{lp_baryon_lfwf}
\end{alignat}
where the dependence of the $ \bar{ n } $-collinear fields on the plus-components of positions is next-to-next-to-leading power and therefore is not relevant for our calculations. The $ n $-directed Wilson lines are defined as
\begin{equation}
\left[ L n + b_{ q }, y_{ q }^{ - } n + b_{ q } \right]_{ \bar{ n } } = P \! \left[ \exp \! \left( i g \int_{ y_{ q }^{ - } }^{ L } \! d \sigma n^{ \mu } A_{ \bar{ n }, \mu } \left( \sigma n + b_{ q } \right) \right) \right] \! ,
\label{nbarcollinear_nwilson}
\end{equation}
where $ L = - \infty, + \infty $ for a hadron in the initial or final state, respectively, and analogously for the gauge links to transverse infinity. The first CF $ \widehat{ C }_{ 1 } $ in Eq. (\ref{lp_baryon_qtmd_correlator_factorized}) acts on the first quark operator, and so on.

For the lattice factor, we have
\begin{alignat}{2}
  \Psi \! \left( b_{ 1 }, b_{ 2 }, b_{ 3 } \right)
& ={}
& \langle 0 \rvert \frac{ 1 }{ 3! } \varepsilon_{ i j k }
& H^{ \dagger }_{ i } \! \left( b_{ 1 } \right) \left[ b_{ 1 }, L \bar{ n } + b_{ 1 } \right]_{ v } \left[ L \bar{ n } + b_{ 1 }, L \bar{ n } + \infty_{ \perp } \right]_{ v, i'' } \nonumber \\
&
& \times
& H^{ \dagger }_{ j } \! \left( b_{ 2 } \right) \left[ b_{ 2 }, L \bar{ n } + b_{ 2 } \right]_{ v } \left[ L \bar{ n } + b_{ 2 }, L \bar{ n } + \infty_{ \perp } \right]_{ v, j'' } \nonumber \\
&
& \times
& H^{ \dagger }_{ k } \! \left( b_{ 3 } \right) \left[ b_{ 3 }, L \bar{ n } + b_{ 3 } \right]_{ v } \left[ L \bar{ n } + b_{ 3 }, L \bar{ n } + \infty_{ \perp } \right]_{ v, k'' } \varepsilon_{ i'' j'' k'' } \lvert 0 \rangle,
\label{baryon_lfwf_lattice_factor}
\end{alignat}
where $ H^{ \dagger } $ is defined in Eq. (\ref{vcollinear_vwilson_line_def}), while the transverse and $ \bar{ n } $-directed Wilson lines are defined analogously to Eq. (\ref{nbarcollinear_nwilson}), with the replacement $ A_{ \bar{ n } } \mapsto A_{ v } $. Throughout, we implicitly contract the fundamental-representation color indices through the Wilson lines. 

The appropriate soft factor is given by
\begin{alignat}{4}
  S \! \left( b_{ 1 }, b_{ 2 }, b_{ 3 } \right) = \langle 0 \rvert \frac{ 1 }{ 3! } \varepsilon_{ i j k } 
& \left[ L n + L_{ \perp }, L n + b_{ 1 } \right]_{ i }
&
& \left[ L n + b_{ 1 }, b_{ 1 } \right]
&
& \left[ b_{ 1 }, L \bar{ n } + b_{ 1 } \right]
&
& \left[ L \bar{ n } + b_{ 1 }, L \bar{ n } + L_{ \perp } \right]_{ i'' } \nonumber \\
  \times
& \left[ L n + L_{ \perp }, L n + b_{ 2 } \right]_{ j }
&
& \left[ L n + b_{ 2 }, b_{ 2 } \right]
&
& \left[ b_{ 2 }, L \bar{ n } + b_{ 2 } \right]
&
& \left[ L \bar{ n } + b_{ 2 }, L \bar{ n } + L_{ \perp } \right]_{ j'' } \nonumber \\
  \times
& \left[ L n + L_{ \perp }, L n + b_{ 3 } \right]_{ k }
&
& \left[ L n + b_{ 3 }, b_{ 3 } \right]
&
& \left[ b_{ 3 }, L \bar{ n } + b_{ 3 } \right]
&
& \left[ L \bar{ n } + b_{ 3 }, L \bar{ n } + L_{ \perp } \right]_{ k'' } \varepsilon_{ i'' j'' k'' } \lvert 0 \rangle .
\label{baryon_lfwf_soft_factor}
\end{alignat}
\end{widetext}

\begin{figure}[ht]
\centering
\includegraphics[scale = 0.83]{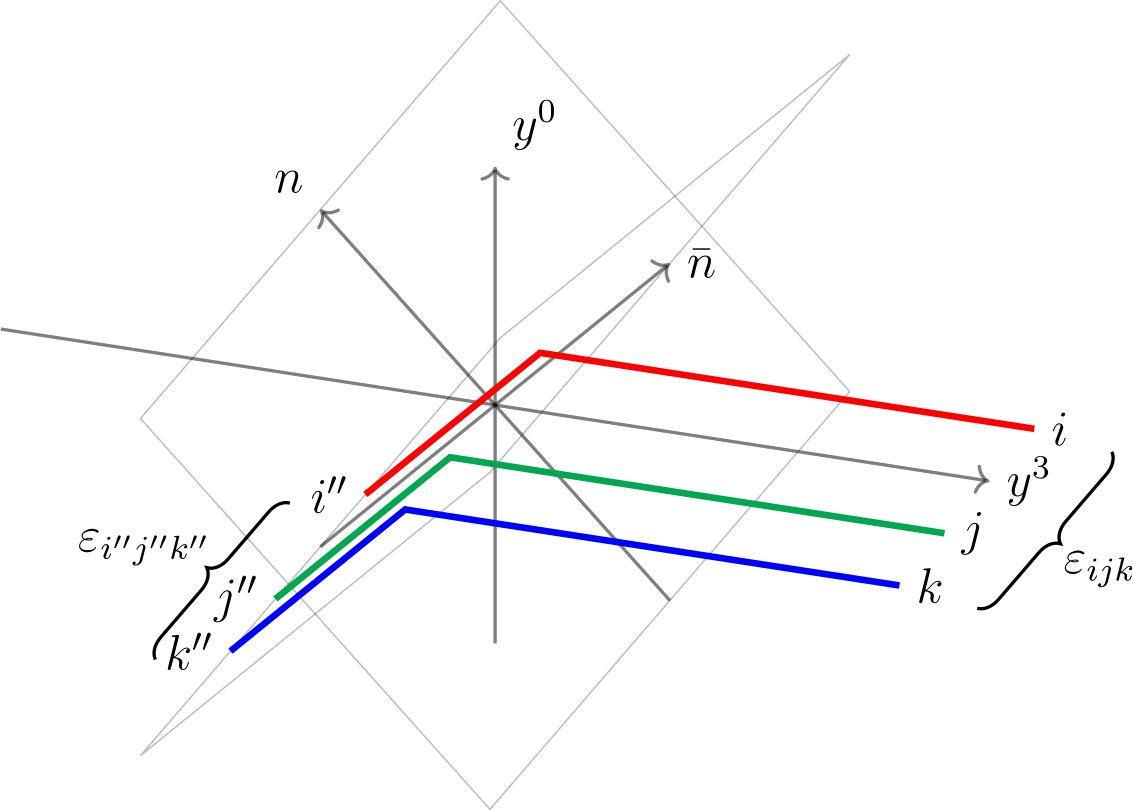}
\caption{Lattice factor for the baryon three-quark LFWF. The lightlike and $ v $-directed Wilson lines connect on the hypersurface $ y^{ 3 } = 0 $. They are colored in red, green and blue to visually indicate the antisymmetrization of the fundamental-representation color indices on both ends. For simplicity, the transverse gauge links on both ends are implied.}
\label{baryon_lfwf_lattice_factor_fig}
\end{figure}

\begin{figure}[ht]
\centering
\includegraphics[scale = 0.83]{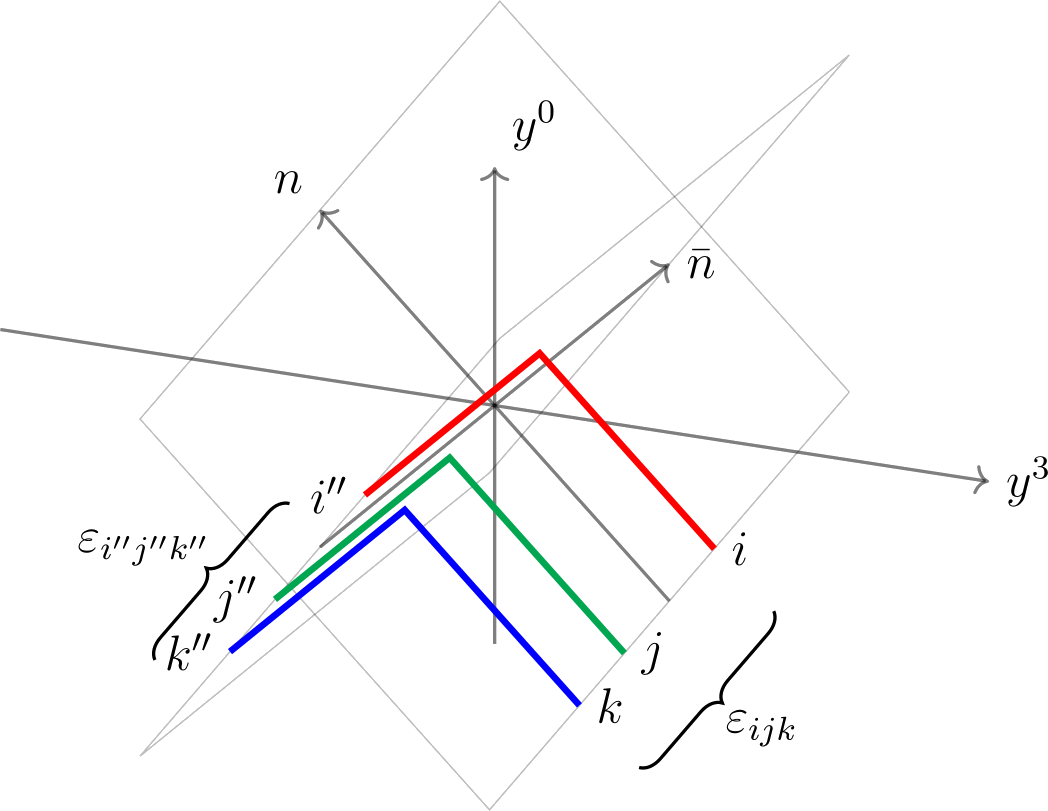}
\caption{Soft factor for the three-quark baryon LFWF. The lightlike Wilson lines orthogonal to $ \bar{ n } $ and $ n $ connect on the hypersurface $ y^{ 3 } = 0 $. They are colored in red, green and blue to visually indicate the antisymmetrization of the fundamental-representation color indices at both ends. For simplicity, the transverse gauge links at both ends are implied.}
\label{baryon_lfwf_soft_factor_fig}
\end{figure}

From now on, we will focus on the case of a baryon in the initial state. The lattice factor and soft factor are represented in Figs. \ref{baryon_lfwf_lattice_factor_fig} and \ref{baryon_lfwf_soft_factor_fig}, respectively.

From Eq. (\ref{lp_nlo_cf_longitudinal_momentum_space}), in longitudinal-momentum-fraction space we have
\begin{align}
& \Omega_{ v; B } \! \left( \left\{ x_{ q }, b_{ q } \right\}_{ q = 1, 2, 3 } \right) \nonumber \\
& = \frac{ \Psi \! \left( \left\{ b_{ q } \right\} \right) C_{ 1 } \! \left( x_{ 1 } \right) C_{ 1 } \! \left( x_{ 2 } \right) C_{ 1 } \! \left( x_{ 3 } \right) \Phi_{ 1 1 1 } \! \left( \left\{ x_{ q }, b_{ q } \right\} \right) }{ S \! \left( \left\{ b_{ q } \right\} \right) } \nonumber \\
& \times \delta \! \left( x_{ 1 } + x_{ 2 } + x_{ 3 } - 1 \right) \! ,
\label{lp_baryon_qtmd_correlator_factorized_longitudinal_momentum_space}
\end{align}
where
\begin{equation}
\Phi_{ 1 1 1 } \! \left( \left\{ x_{ q }, b_{ q } \right\} \right) = \int \! d^{ 3 } y_{ q }^{ - } e^{ i P^{ + } \sum_{ q } y_{ q }^{ - } x_{ q } } \widetilde{ \Phi }_{ 1 1 1 } \! \left( \left\{ y_{ q }^{ - }, b_{ q } \right\} \right) \! .
\label{lp_baryon_lfwf_longitudinal_momentum_space}
\end{equation}
The $ x_{ q } $ are the fractions of longitudinal momentum of the baryon carried by the quarks, therefore, by momentum conservation, we must have $ \sum_{ q } x_{ q } = 1 $.

Going beyond leading order, we encounter an interconnected structure of divergences. Alongside ultraviolet (UV) and infrared (IR) divergences, we have rapidity divergences, when a lightlike component of a loop momentum $ k $ goes to infinity at constant $ k^{ 2 } $. The UV and IR divergences are regulated using dimensional regularization in $ D = 4 - 2 \epsilon $ dimensions, while rapidity divergences we handled using $ \delta $-regularization, i.e.,
\begin{align}
  \left[ L n + z, z \right]
& \mapsto P \! \left[ \exp \! \left( i g \int_{ 0 }^{ L } \! d \sigma n^{ \mu } A_{ \mu } \! \left( \sigma n + z \right) e^{ - \lvert \sigma \rvert \delta^{ + } } \right) \right] \! ,
\label{delta_regularization_n_direction} \\
  \left[ z, L \bar{ n } + z \right]
& \mapsto P \! \left[ \exp \! \left( i g \int_{ L }^{ 0 } \! d \chi \bar{ n }^{ \mu } A_{ \mu } \! \left( \chi \bar{ n } + z \right) e^{ - \lvert \chi \rvert \delta^{ - } } \right) \right] \! ,
\label{delta_regularization_nbar_direction}
\end{align}
where $ \delta^{ \pm } > 0 $ have dimensions of mass, and we define $ \delta^{ 2 } = 2 \delta^{ + } \delta^{ - } $. Rapidity divergences are multiplicatively renormalizable, as we will see explicitly at next-to-leading order, after regularization and the introduction of momentum scales $ \nu_{ q }^{ + }, \nu_{ q }^{ - } $ for each quark and for direction $ n, \bar{ n } $, respectively, with $ \nu_{ q }^{ 2 } = 2 \nu_{ q }^{ + } \nu_{ q }^{ - } $.

The divergences from the factorization procedure must recombine so that only the ones of the original QTMD correlator remain, and, up to next-to-leading power, all divergences must cancel independently for each leg. Moreover, in order to extract a physical LFWF, it must be renormalizable independently of the lattice factor. We will verify all of this in the following section at next-to-leading order, while in the remainder of this section we calculate all the divergent contributions, working in momentum space and choosing the $ \overline{ \text{MS} } $ renormalization scheme. Note that the calculations proceed identically for any number of colors $ N $, provided we consider $ N $-point operators. To avoid unnecessary notational clutter, we focus on the three-quark case of primary interest, but we evaluate the color structures for generic $ N $ once and for all.

\subsection{Divergences of the Light-Front Wave Function}
\label{lfwf_divergences_subsec}

We start by calculating the divergences of the three-quark LFWF (\ref{lp_baryon_lfwf}) at next-to-leading order. They come from one-loop interactions between a quark and a lightlike Wilson line, either the one directly connected to it, or the gauge link associated to another quark.

\begin{figure}[ht]
\centering
\includegraphics[scale = 1]{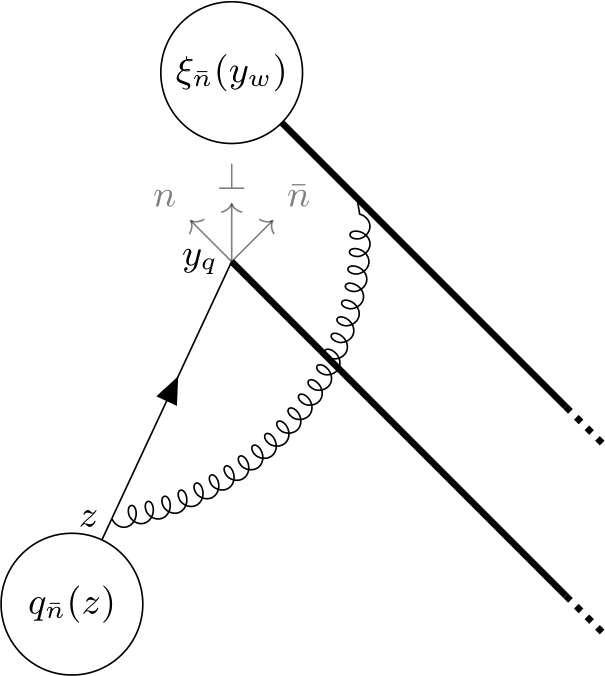}
\caption{One-loop diagram for the interaction between a quark and a $ n $-directed Wilson line at different transverse positions.}
\label{quark_nwilson_oneloop_with_transversep_fig}
\end{figure}

We first consider the divergences originating from interactions with a transverse separation, diagrammatically represented in Fig.~\ref{quark_nwilson_oneloop_with_transversep_fig}.
Going back to the TMD expansion with two background fields, and once again choosing the Feynman gauge for dynamical fields and the LC gauge (\ref{lc_gauge_for_bkg_fields}), (\ref{residual_lc_gauge_for_bkg_fields}) for the background fields, we can construct the contributions from the analogous of the contraction in (\ref{lp_nlo_effective_quark_current_contraction}). For, e.g., quark 1 interacting with the Wilson line of quark 2, we have to consider the contraction
\begin{align}
& \varepsilon_{ i j k } \delta_{ i i' } \delta_{ k k' } \xi_{ \bar{ n }, j' } \! \left( y_{ 2 } \right) \xi_{ \bar{ n }, k' } \nonumber \\
& \times i g \wick{ \int_{ y_{ 2 }^{ - } }^{ L } \! d \sigma n^{ \mu } \c3{ B }_{ \mu, j j' } \! \left( \sigma n + b_{ 2 } \right) \c2{ \psi }_{ i' } \! \left( y_{ 1 } \right) i g \int \! d^{ D } z \c2{ \overline{ \psi } } \c3{ \slashed{ B } } q_{\bar{ n }  } \! \left( z \right) \! , }
\label{nlo_lp_baryon_lfwf_12_contraction}
\end{align}
where $ \delta_{ i i' } $ and $ \delta_{ k k' } $ come from the order-zero expansions of the dynamical Wilson lines for quark 1 and quark 3. Using Eqs. (\ref{dirac_propagator_position_space}), (\ref{gauge_propagator_position_space}), for the color structure we have
\begin{align}
& \varepsilon_{ i j k } \left( \sum_{ A = 1 }^{ N^{ 2 } - 1 } t^{ A }_{ j j' } t^{ A }_{ i l} \right) q_{ \bar{ n }, l } \! \left( z \right) \xi_{ \bar{ n }, j' } \! \left( y_{ 2 } \right) \xi_{ \bar{ n }, k } \! \left( y_{ 3 } \right) \nonumber \\
& = - \frac{ C_{ F } }{ N - 1 } \varepsilon_{ i j k } q_{ \bar{ n }, i } \! \left( z \right) \xi_{ \bar{ n }, j } \! \left( y_{ 2 } \right) \xi_{ \bar{ n }, k } \! \left( y_{ 3 } \right) \! ,
\label{nlo_lp_baryon_lfwf_12_color_struct}
\end{align}
where we used Eq. (\ref{product_of_generators_of_sun}), and $ C_{ F } $ is defined in Eq. (\ref{cf}). Projecting with $ \Lambda_{ + } $ as in Eq. (\ref{lc_good_projector}) to retain only LP contributions, we have
\begin{widetext}
\begin{equation}
\text{diag}_{ 1 2 }  = - \! \frac{ i g^{ 2 } }{ 8 \pi^{ D } } \frac{ C_{ F } }{ N - 1 } \Gamma \! \left( 1 - \epsilon \right) \Gamma \! \left( 2 - \epsilon \right) \int_{ y_{ 2 }^{ - } }^{ L } \! d \sigma \int \! d^{ D } z \varepsilon_{ i j k } \frac{ \Lambda_{+} \left( y_{ 1 }^{ - } \gamma^{ + } + \slashed{ b_{ 1 } } - \slashed{ z } \right) \gamma^{ + } q_{ \bar{ n }, i } \! \left( z \right) \xi_{ \bar{ n }, j } \! \left( y_{ 2 } \right) \xi_{ \bar{ n }, k } \! \left( y_{ 3 } \right) }{ \left( - \! \left( \sigma n + b_{ 2 } - z \right)^{ 2 } + i 0 \right)^{ 1 - \epsilon } \left( - \left( y_{ 1 }^{ - } n + b_{ 1 } - z \right)^{ 2 } + i 0 \right)^{ 2 - \epsilon } }.
\label{nlo_lp_baryon_lfwf_12_contribution}
\end{equation}
\end{widetext}
After integrating, and keeping only the rapidity divergent part, we find
\begin{alignat}{1}
  \text{diag}_{ 1 2 } ={}
& - 2 a_{ s } \frac{ C_{ F } }{ N - 1 } \Gamma \! \left( - \epsilon \right) \left( \frac{ - b_{ 2 1 }^{ 2 }}{ 4 } \right)^{ \! \epsilon } \ln \! \left( \frac{ \delta^{ + } }{ i s \hat{ p }_{ 1 }^{ + } } \right) \nonumber \\
& \times \varepsilon_{ i j k } \xi_{ \bar{ n }, i } \! \left( y_{ 1 } \right) \xi_{ \bar{ n }, j } \! \left( y_{ 2 } \right) \xi_{ \bar{ n }, k } \! \left( y_{ 3 } \right) + ...,
\label{nlo_lp_baryon_lfwf_12_contribution_rapidity_finite_ignored}
\end{alignat}
where the ellipsis represent rapidity-finite terms. Note that this contribution, as well as all the other rapidity-finite terms not explicitly considered, are also IR divergent, as evident from the factor of $ \Gamma \! \left( - \epsilon \right) $. These divergences can be ignored for the purpose of studying the scale evolution of the LFWF, therefore we also remove the pole at $ \epsilon = 0 $ from Eq. (\ref{nlo_lp_baryon_lfwf_12_contribution_rapidity_finite_ignored}). The contributions from quark 2 interacting with the Wilson line of quark 1, and all other combinations, are calculated in exactly the same way. Therefore, the total NLO rapidity-divergent contribution is
\begin{widetext}
\begin{equation}
\sum_{ \substack{ q, w = 1 \\ w \neq q } }^{ N } \text{diag}_{ q w } = - 2 a_{ s } \frac{ C_{ F } }{ N - 1 } \sum_{ \substack{ q, w = 1 \\ w \neq q } }^{ N } \left( \Gamma \! \left( - \epsilon \right) \left( \frac{ - \left( b_{ w } - b_{ q } \right)^{ 2 } }{ 4 } \right)^{ \! \epsilon } + \frac{ 1 }{ \epsilon } \right) \ln \! \left( \frac{ \delta^{ + } }{ i \text{sign} \! \left( L \right) \hat{ p }_{ q }^{ + } } \right) \varepsilon_{ i j k } \xi_{ \bar{ n }, i } \! \left( y_{ 1 } \right) \xi_{ \bar{ n }, j } \! \left( y_{ 2 } \right) \xi_{ \bar{ n }, k } \! \left( y_{ 3 } \right) \! .
\label{nlo_lp_baryon_lfwf_total_rapidity_diverget_contribution}
\end{equation}
\end{widetext}

\begin{figure}[ht]
\centering
\includegraphics[scale = 1]{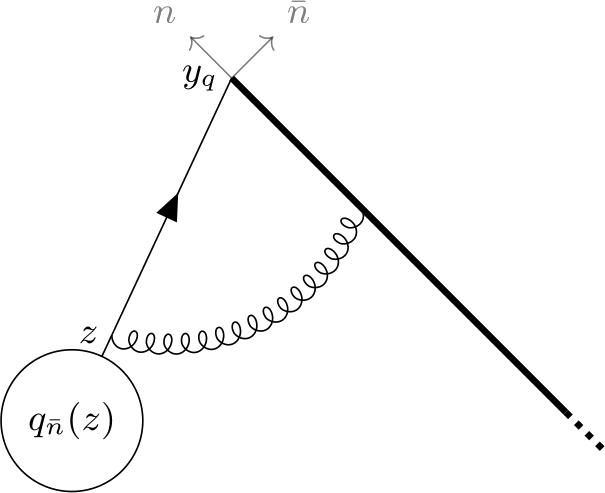}
\caption{One-loop diagram for the interaction between a quark and a $ n $-directed Wilson line at the same transverse position.}
\label{quark_nwilson_oneloop_no_transversep_fig}
\end{figure}

Next, we consider the NLO contributions from a gluon exchange between a quark and the corresponding Wilson line, diagrammatically represented in Fig \ref{quark_nwilson_oneloop_no_transversep_fig}.
The integral to be evaluated is analogous to the case with transverse separation, but its absence makes it UV divergent. The calculation proceeds in a similar fashion to the previous one, but it generates a different color structure. Considering, e.g., quark 1, we have
\begin{align}
& \varepsilon_{ i j k } \left( \sum_{ A = 1 }^{ N^{ 2 } - 1 } t^{ A }_{ i i' } t^{ A }_{ i' l} \right) q_{ \bar{ n }, l } \! \left( z \right) \xi_{ \bar{ n }, j } \! \left( y_{ 2 } \right) \xi_{ \bar{ n }, k } \! \left( y_{ 3 } \right) \nonumber \\
& = C_{ F } \varepsilon_{ i j k } q_{ \bar{ n }, i } \! \left( z \right) \xi_{ \bar{ n }, j } \! \left( y_{ 2 } \right) \xi_{ \bar{ n }, k } \! \left( y_{ 3 } \right) \! .
\label{nlo_lp_baryon_lfwf_11_color_struct}
\end{align}
Adapting Eq. (\ref{nlo_lp_baryon_lfwf_12_contribution_rapidity_finite_ignored}), we find
\begin{alignat}{1}
  \text{diag}_{ 1 1 } =
& + 2 a_{ s } C_{ F } \Gamma \! \left( \epsilon \right) \left( 1 + \ln \! \left( \frac{ \delta^{ + } }{ i \text{sign} \! \left( L \right) \hat{ p }_{ 1 }^{ + } } \right) \right) \nonumber \\
& \times \varepsilon_{ i j k } \xi_{ \bar{ n }, i } \! \left( y_{ 1 } \right) \xi_{ \bar{ n }, j } \! \left( y_{ 2 } \right) \xi_{ \bar{ n }, k } \! \left( y_{ 3 } \right) \! ,
\label{nlo_lp_baryon_lfwf_total_uv_diverget_contribution}
\end{alignat}
where $ \delta^{ + } $ now regularizes the collinear divergences from the interaction with the Wilson line at infinity. These are conceptually different from rapidity divergences, since they do not depend on the transverse separation. The calculations proceed identically for each quark, therefore we can extract the UV renormalization constant for the LP field up to next-to-leading order. In momentum space, and in the $ \overline{ \text{MS} } $ renormalization scheme, we have
\begin{align}
& \widetilde{ Z }_{ \Phi 1; \overline{ \text{MS} } } \! \left( \frac{ \delta^{ + } }{ p_{ q }^{ + } } \right) = 1 + 2 a_{ s } C_{ F } \Gamma \! \left( \epsilon \right) e^{ \epsilon \gamma_{ E } } \left( \mu^{ 2 } \right)^{ \epsilon } \nonumber \\
& \times \left( 1 + \ln \! \left( \frac{ \delta^{ + } }{ i \text{sign} \! \left( L \right) p_{ q }^{ + } } \right) \right) + \mathcal{ O } \! \left( a_{ s }^{ 2 } \right) \nonumber \\
& = 1 + 2 a_{ s } C_{ F } \frac{ 1 }{ \epsilon } \left( 1 + \ln \! \left( \frac{ \delta^{ + } }{ i \text{sign} \! \left( L \right) p_{ q }^{ + } } \right) \right) + \mathcal{ O } \! \left( a_{ s }^{ 2 } \right) \! ,
\label{lp_quark_uv_renormalization_constant_nlo_explicit_vertex_only}
\end{align}
where $ p_{ q }^{ + } $ is the longitudinal momentum of quark $ q $, and in the second equality we expanded around $ \epsilon = 0 $. Incorporating the quark self-energy, i.e.~\cite{Vladimirov:2021hdn},
\begin{equation}
Z_{ 2; \overline{ \text{MS} } }^{ \frac{ 1 }{ 2 } } = 1 - a_{ s } C_{ F } \frac{ 1 }{ 2 \epsilon } + \mathcal{ O } \! \left( a_{ s }^{ 2 } \right) \! ,
\label{quark_renormalization_constant_nlo_explicit}
\end{equation}
we have
\begin{alignat}{1}
  Z_{ \Phi 1; \overline{ \text{MS} } } \! \left( \frac{ \delta^{ + } }{ p_{ q }^{ + } } \right) = 1
& + a_{ s } C_{ F } \frac{ 1 }{ \epsilon } \left( \frac{ 3 }{ 2 } + 2 \ln \! \left( \frac{ \delta^{ + } }{ i \text{sign} \! \left( L \right) p_{ q }^{ + } } \right) \right) \nonumber \\
& + \mathcal{ O } \! \left( a_{ s }^{ 2 } \right) \! .
\label{lp_quark_uv_renormalization_constant_nlo_explicit}
\end{alignat}
Note that the self-energy of an $ n $-directed Wilson line is zero, since it is calculated from the contraction of two terms $ \sim n^{ \mu } B_{ \mu } $, and therefore it is proportional to $  n^{ 2 } = 0 $.

By moving to momentum space and introducing momentum scales $ \nu_{ q }^{ + } $, we extract from Eq. (\ref{nlo_lp_baryon_lfwf_total_rapidity_diverget_contribution}) the correction from rapidity divergences to the LP LFWF up to next-to-leading order. In the $ \overline{ \text{MS} } $ scheme, we have
\begin{equation}
R_{ \overline{ \text{MS} } } \! \left( \delta^{ + }, \left\{ \nu_{ q }^{ + } \right\} \right) = \prod_{ q = 1 }^{ N } R_{ \overline{ \text{MS} } } \! \left( \frac{ \delta^{ + } }{ \nu_{ q }^{ + } } \right) \! ,
\label{lp_baryon_lfwf_rapidity_divergent_correction_nlo_explicit}
\end{equation}
where the correction from each quark is given by
\begin{alignat}{1}
  R_{ \overline{ \text{MS} } } \! \left( \frac{ \delta^{ + } }{ \nu_{ q }^{ + } } \right) = 1
& + 2 a_{ s } \frac{ C_{ F } }{ N - 1 } \sum_{ \substack{ w = 1 \\ w \neq q } }^{ N } \ln \! \left( \frac{ - \left( b_{ w } - b_{ q } \right)^{ 2 } }{ 4 } \mu^{ 2 } \right) \nonumber \\
& \times \ln \! \left( \frac{ \delta^{ + } }{ \nu_{ q }^{ + } } \right) + \mathcal{ O } \! \left( a_{ s }^{ 2 } \right) \! .
\label{lp_quark_rapidity_divergent_correction_nlo_explicit}
\end{alignat}
Note that we dropped the nondivergent terms $ \sim \ln \! \left( - \mu^{ 2 } \left( b_{ w } - b_{ q } \right)^{ 2 } / 4 \right) \ln \! \left( \nu_{ q }^{ + } / p_{ q }^{ + } \right) $.

The results are equivalent to those for TMD parton distribution functions and fragmentation functions in Ref.~\cite{Vladimirov:2021hdn}, taking into account the different color structure, and the fact that our Wilson lines always extend toward infinity. Since there is no contraction between Dirac adjoints, the imaginary parts of the renormalization constants do not compensate, and therefore we kept their contribution in Eq. (\ref{lp_quark_uv_renormalization_constant_nlo_explicit}).

Analogous results would be found in any gauge. In a gauge-invariant formulation, alongside the contraction (\ref{nlo_lp_baryon_lfwf_12_contraction}), we would have to consider all other terms in the expansion of the Wilson line with exactly one dynamical field, as well as the terms in the remaining two links with no dynamical fields. Ultimately, we are left with the gauge links in the background. Two quarks are connected to the Levi-Civita symbol through their background Wilson lines, allowing us to directly adapt the procedure used previously. To understand how we recover the last background link, consider, e.g., the three contractions from the $ \mathcal{ O } \! \left( g^{ 3 } \right) $-expansion, using only a single type of background field for simplicity. We have
\begin{alignat}{2}
&
&
& + \int_{ a }^{ b } \! d x_{ 1 } A \! \left( x_{ 1 } \right) \int_{ x_{ 1 } }^{ b } \! d x_{ 2 } A \! \left( x_{ 2 } \right) \int_{ x_{ 2 } }^{ b } \! d x_{ 3 } B \! \left( x_{ 3 } \right) \nonumber \\
&
&
& + \int_{ a }^{ b } \! d x_{ 1 } A \! \left( x_{ 1 } \right) \int_{ x_{ 1 } }^{ b } \! d x_{ 2 } B \! \left( x_{ 2 } \right) \int_{ x_{ 2 } }^{ b } \! d x_{ 3 } A \! \left( x_{ 3 } \right) \nonumber \\
&
&
& + \int_{ a }^{ b } \! d x_{ 1 } B \! \left( x_{ 1 } \right) \int_{ x_{ 1 } }^{ b } \! d x_{ 2 } A \! \left( x_{ 2 } \right) \int_{ x_{ 2 } }^{ b } \! d x_{ 3 } A \! \left( x_{ 3 } \right) \nonumber \\
& =
&
& + \int_{ a }^{ b } \! d x_{ 1 } A \! \left( x_{ 1 } \right) \int_{ x_{ 1 } }^{ b } \! d x_{ 2 } A \! \left( x_{ 2 } \right) \int_{ x_{ 2 } }^{ b } \! d x_{ 3 } B \! \left( x_{ 3 } \right) \nonumber \\
&
&
&  + \int_{ a }^{ b } \! d x_{ 1 } A \! \left( x_{ 1 } \right) \int_{ x_{ 1 } }^{ b } \! d x_{ 3 } A \! \left( x_{ 3 } \right) \int_{ x_{ 1 } }^{ x_{ 3 } } \! d x_{ 2 } B \! \left( x_{ 2 } \right) \nonumber \\
&
&
& + \int_{ a }^{ b } \! d x_{ 2 } A \! \left( x_{ 2 } \right) \int_{ a }^{ x_{ 2 } } \! d x_{ 1 } B \! \left( x_{ 1 } \right) \int_{ x_{ 2 } }^{ b } \! d x_{ 3 } A \! \left( x_{ 3 } \right) \nonumber \\
& =
&
& + \int_{ a }^{ b } \! d x_{ 1 } A \! \left( x_{ 1 } \right) \int_{ x_{ 1 } }^{ b } \! d x_{ 2 } A \! \left( x_{ 2 } \right) \int_{ x_{ 2 } }^{ b } \! d x_{ 3 } B \! \left( x_{ 3 } \right) \nonumber \\
&
&
&  + \int_{ a }^{ b } \! d x_{ 1 } A \! \left( x_{ 1 } \right) \int_{ x_{ 1 } }^{ b } \! d x_{ 2 } A \! \left( x_{ 2 } \right) \int_{ x_{ 1 } }^{ x_{ 2 } } \! d x_{ 3 } B \! \left( x_{ 3 } \right) \nonumber \\
&
&
& + \int_{ a }^{ b } \! d x_{ 1 } A \! \left( x_{ 1 } \right) \int_{ a }^{ x_{ 1 } } \! d x_{ 3 } B \! \left( x_{ 3 } \right) \int_{ x_{ 1 } }^{ b } \! d x_{ 2 } A \! \left( x_{ 2 } \right) \nonumber \\
& =
&
& + \int_{ a }^{ b } \! d x_{ 1 } A \! \left( x_{ 1 } \right) \int_{ x_{ 1 } }^{ b } \! d x_{ 2 } A \! \left( x_{ 2 } \right) \int_{ a }^{ b } \! d x_{ 3 } B \! \left( x_{ 3 } \right) \! ,
\label{eg_gcubed_wilson_expansion_integrals}
\end{alignat}
where in the second equality we renamed the integration variables so that $ x_{ 2 } \leftrightarrow x_{ 3 } $ in the second term, and $ x_{ 2 } \rightarrow x_{ 1 } \rightarrow x_{ 3 } \rightarrow x_{ 2 } $ in the third term. Considering all terms in the expansion of the Wilson line with one dynamical field, we recover that the exchanged gluon can attach to any point along the path of the gauge link, while the last quark remains connected to the Levi-Civita symbol through a background Wilson line.

\subsection{Divergences of the Lattice Factor}
\label{lattice_divergences_subsec}

We now turn to the NLO divergent contributions to the lattice factor (\ref{baryon_lfwf_lattice_factor}), which come from one-gluon exchanges between a $ v $-directed and a $ \bar{ n } $-directed Wilson line.

\begin{figure}[ht]
\centering
\includegraphics[scale = 1]{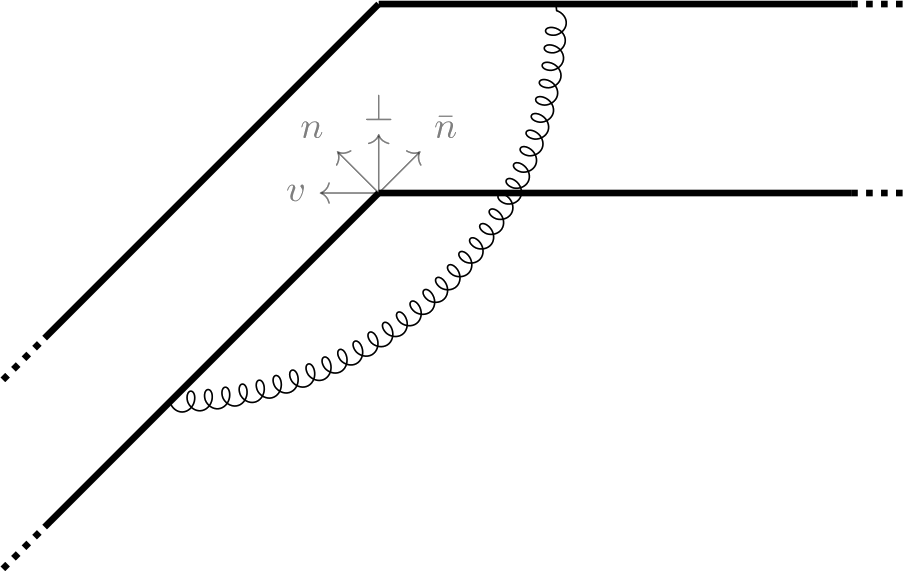}
\caption{One-loop diagram for a gluon exchange between a $ v $ and a $ \bar{ n } $-directed Wilson line at different transverse positions.}
\label{vilson_nbarwilson_oneloop_with_transversep_fig}
\end{figure}

First, we consider the divergences originating from gluon exchanges with a transverse separation, diagrammatically represented in Fig.~\ref{vilson_nbarwilson_oneloop_with_transversep_fig}.
For, e.g., gauge links at $ b_{ 1 } $ and $ b_{ 2 } $, we have to consider the contraction
\begin{align}
& \varepsilon_{ i j k } \delta_{ i i' } \delta_{ k k'' } \delta_{ j' j'' } \varepsilon_{ i'' j'' k'' } v^{ \mu } \nonumber \\
& \times i g \wick{ \int_{ 0 }^{ L } \! d \sigma \c3{ B }_{ \mu, i' i'' } \! \left( \sigma v + b_{ 1 } \right) i g \int_{ L }^{ 0 } \! d \chi \bar{ n }^{ \nu } \c3{ B }_{ \nu, j j' } \! \left( \chi \bar{ n } + b_{ 2 } \right) \! , }
\label{nlo_baryon_lfwf_lattice_12_contraction}
\end{align}
where the Kronecker delta come from the order-zero expansions of the remaining Wilson lines. Using Eq. (\ref{gauge_propagator_position_space}), for the color structure we have
\begin{equation}
\frac{ 1 }{ N! } \varepsilon_{ i j k } \left( \sum_{ A = 1 }^{ N^{ 2 } - 1 } t^{ A }_{ i i'' } t^{ A }_{ j j'' } \right) \varepsilon_{ i'' j'' k } = - \frac{ C_{ F } }{ N - 1 },
\label{nlo_baryon_lfwf_lattice_12_color_struct}
\end{equation}
where we used Eq. (\ref{product_of_generators_of_sun}). Therefore, defining $ s = \text{sign} \! \left( L \right) $ and including $ \delta $-regularization, we have
\begin{alignat}{1}
  \text{diag}_{ 1 2 }' ={}
& - \! \frac{ g^{ 2 } }{ 4 \pi^{ 2 - \epsilon } } \frac{ C_{ F } }{ N - 1 } \Gamma \! \left( 1 - \epsilon \right) \int_{ 0 }^{ L } \! d \sigma \int_{ L }^{ 0 } \! d \chi v \bar{ n } e^{ -  \delta^{ - } s \chi } \nonumber \\
& \times \frac{ 1 }{ \left( - \! \left( \sigma v + b_{ 1 } - \chi \bar{ n } - b_{ 2 } \right)^{ 2 } + i 0 \right)^{ 1 - \epsilon } } \Psi .
\label{nlo_baryon_lfwf_lattice_12_contribution}
\end{alignat}
Similarly to the LFWF, we isolate the rapidity divergent contributions, alongside those from the other combinations of Wilson lines. The total NLO contribution from diagrams with transverse separation is
\begin{align}
& \sum_{ \substack{ q, w = 1 \\ w \neq q } }^{ N } \text{diag}_{ q w }' = + 2 a_{ s } C_{ F } \sum_{ q = 1 }^{ N } \Gamma \! \left( 1 - 2 \epsilon \right) \Gamma \! \left( 2 \epsilon \right) \Gamma \! \left( \epsilon \right) \nonumber \\
& \times \left( \frac{ \text{sign} \! \left( L \right) \sqrt{ - v^{ 2 } } \delta^{ - } }{  v^{ - } } \right)^{ \! - 2 \epsilon } \Psi \nonumber \\
& - a_{ s } \frac{ C_{ F } }{ N - 1 } \sum_{ \substack{ q, w = 1 \\ w \neq q } }^{ N } \Gamma \! \left( - \epsilon \right) \left( \frac{ - \left( b_{ q } - b_{ w } \right)^{ 2 } }{ 4 } \right)^{ \! \epsilon } \nonumber \\
& \times \left( - \psi \! \left( - \epsilon \right)  - \gamma_{ E } + \ln \! \left( \frac{ - \left( b_{ q } - b_{ w } \right)^{ 2 } v^{ 2 } }{ 4 } \left( \frac{ \delta^{ - } }{ v^{ - } } \right)^{ \! 2 } \right) \right) \Psi ,
\label{nlo_baryon_lfwf_lattice_total_contribution_with_transversep}
\end{align}
where $ \psi $ is the Digamma function, i.e., $ \psi \! \left( z \right) = \left( \Gamma' / \Gamma \right) \! \left( z \right) $.

\begin{figure}[ht]
\centering
\includegraphics[scale = 1]{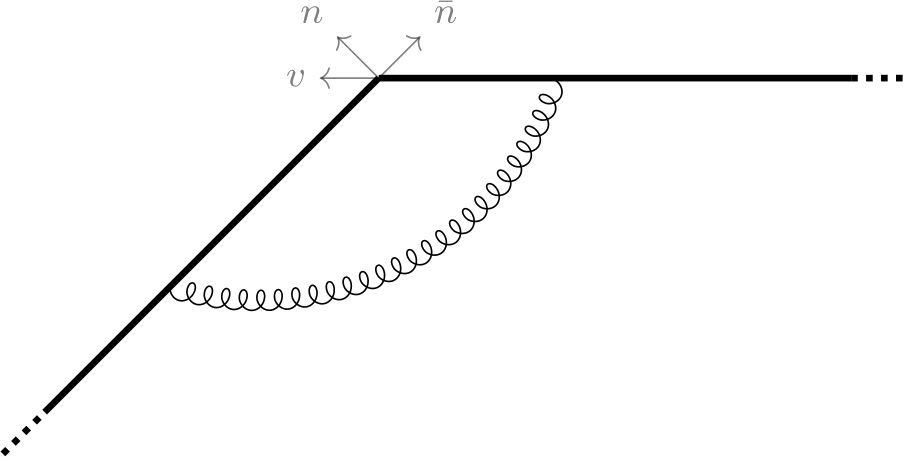}
\caption{One-loop diagram for a gluon exchange between a $ v $ and a $ \bar{ n } $-directed Wilson line at the same transverse position.}
\label{vilson_nbarwilson_oneloop_no_transversep_fig}
\end{figure}

Next, we consider the NLO contributions from a gluon exchange between a $ v $-directed Wilson line and the corresponding lightlike link, diagrammatically represented in Fig. \ref{vilson_nbarwilson_oneloop_no_transversep_fig}.
The calculations are analogous to the case with transverse separation. The only difference is in the color structure, e.g., for the Wilson lines at $ b_{ 1 } $ we have
\begin{equation}
\frac{ 1 }{ N! } \varepsilon_{ i j k } \delta_{ j j'' } \delta_{ k k'' } \left( \sum_{ A = 1 }^{ N^{ 2 } - 1 } t^{ A }_{ i i' } t^{ A }_{ i' i'' } \right) \varepsilon_{ i'' j'' k'' } = C_{ F }.
\label{nlo_baryon_lfwf_lattice_11_color_struct}
\end{equation}
Adapting Eq. (\ref{nlo_baryon_lfwf_lattice_total_contribution_with_transversep}), for the total NLO contribution from diagrams with zero transverse separation we find
\begin{alignat}{1}
  \sum_{ q = 1 }^{ N } \text{diag}_{ q q }' ={}
& - 2 a_{ s } C_{ F } \sum_{ q = 1 }^{ N } \Gamma \! \left( 1 - 2 \epsilon \right) \Gamma \! \left( 2 \epsilon \right) \Gamma \! \left( \epsilon \right) \nonumber \\
& \times \left( \frac{ \text{sign} \! \left( L \right) \sqrt{ - v^{ 2 } } \delta^{ - } }{  v^{ - } } \right)^{ \! - 2 \epsilon } \Psi.
\label{nlo_baryon_lfwf_lattice_total_contribution_no_transversep}
\end{alignat}
Adding up Eqs. (\ref{nlo_baryon_lfwf_lattice_total_contribution_with_transversep}) and (\ref{nlo_baryon_lfwf_lattice_total_contribution_no_transversep}), there is a compensation between the latter and the first term of the former. The same cancellation occurs for each leg of the lattice factor. Therefore, the total divergent correction in the $ \overline{ \text{MS} } $ renormalization scheme up to next-to-leading order is
\begin{align}
& \widetilde{ \Delta }_{ \Psi; \overline{ \text{MS} } } = 1 - a_{ s } \frac{ C_{ F } }{ N - 1 } \sum_{ \substack{ q, w = 1 \\ w \neq q } }^{ N } \Gamma \! \left( - \epsilon \right) e^{ - \epsilon \gamma_{ E } } \nonumber \\
& \times \left( - \psi \! \left( - \epsilon \right)  - \gamma_{ E } + \ln \! \left( \frac{ - \left( b_{ q } - b_{ w } \right)^{ 2 } }{ 4 } \left( - v^{ 2 } \right) \left( \frac{ \delta^{ - } }{ v^{ - } } \right)^{ \! 2 } \right) \right) \nonumber \\
& \times \left( \frac{ - \left( b_{ q } - b_{ w } \right)^{ 2 } \mu^{ 2 } }{ 4 } \right)^{ \! \epsilon } + \mathcal{ O } \! \left( a_{ s }^{ 2 } \right) \nonumber \\
& =  1 - N a_{ s } C_{ F } \left( \frac{ 1 }{ \epsilon^{ 2 } } - \frac{ 1 }{ \epsilon } \ln \! \left( \frac{ \sqrt{ - v^{ 2 } } \delta^{ - } }{ \mu v^{ - } } \right)^{ \! 2 } \right) \nonumber \\
& + 2 a_{ s } \frac{ C_{ F } }{ N - 1 } \sum_{ \substack{ q, w = 1 \\ w \neq q } }^{ N } \left( \frac{ 1 }{ \epsilon } + \ln \! \left( \frac{ - \left( b_{ q } - b_{ w } \right)^{ 2 } }{ 4 } \mu^{ 2 } \right) \right) \nonumber \\
& \times \ln \! \left( \frac{ \delta^{ - } }{ \nu_{ q }^{ - } } \right) + \mathcal{ O } \! \left( a_{ s }^{ 2 } \right) \nonumber \\
& \coloneq \widetilde{ Z }_{ \Psi 1; \overline{ \text{MS} } }^{ N } \! \left( \frac{ \sqrt{ - v^{ 2 } } \delta^{ - } }{ \mu v^{ - } } \right) R_{ \overline{ \text{MS} } } \! \left( \delta^{ - }, \left\{ \nu_{ q }^{ - } \right\} \right) \! .
\label{nlo_baryon_lfwf_lattice_total_divergent_contribution}
\end{align}
In the second equality, we introduced the momentum scales $ \nu_{ q }^{ - } $, and kept only divergent terms. In the last equality, we defined the renormalization constant
\begin{align}
& \widetilde{ Z }_{ \Psi 1; \overline{ \text{MS} } } \! \left( \frac{ \sqrt{ - v^{ 2 } } \delta^{ - } }{ \mu v^{ - } } \right) = 1 - \frac{ a_{ s } C_{ F } }{ \epsilon^{ 2 } } \nonumber \\
& + \frac{ a_{ s } C_{ F } }{ \epsilon } \ln \! \left( \frac{ \sqrt{ - v^{ 2 } } \delta^{ - } }{ \mu v^{ - } } \right)^{ \! 2 } + \mathcal{ O } \! \left( a_{ s }^{ 2 } \right) \! ,
\label{baryon_lfwf_lattice_uv_renormalization_constant_nlo_explicit_vertex_only}
\end{align}
and the correction from rapidity divergences
\begin{equation}
R_{ \overline{ \text{MS} } } \! \left( \delta^{ - }, \left\{ \nu_{ q }^{ - } \right\} \right) = \prod_{ q = 1 }^{ N } R_{ \overline{ \text{MS} } } \! \left( \frac{ \delta^{ - } }{ \nu_{ q }^{ - } } \right) \! ,
\label{baryon_lfwf_lattice_rapidity_divergent_correction_nlo_explicit}
\end{equation}
with
\begin{alignat}{1}
  R_{ \overline{ \text{MS} } } \! \left( \frac{ \delta^{ - } }{ \nu_{ q }^{ - } } \right) = 1
& + 2 a_{ s } \frac{ C_{ F } }{ N - 1 } \sum_{ \substack{ w = 1 \\ w \neq q } }^{ N } \ln \! \left( \frac{ - \left( b_{ q } - b_{ w } \right)^{ 2 } }{ 4 } \mu^{ 2 } \right) \nonumber \\
& \times \ln \! \left( \frac{ \delta^{ - } }{ \nu_{ q }^{ - } } \right) + \mathcal{ O } \! \left( a_{ s }^{ 2 } \right) \! .
\label{baryon_lfwf_lattice_leg_rapidity_divergent_correction_nlo_explicit}
\end{alignat}
To get the full renormalization constant, we need to add the self-energy contributions of the Wilson lines, which come from contracting the second-order terms in the expansion of the path-ordered exponentials. For $ v $-directed links, in the $ \overline{ \text{MS} } $ renormalization scheme and explicitly up to next-to-leading order, we have~\cite{Rodini:2022wic}
\begin{equation}
Z_{ H; \overline{ \text{MS} } }^{ \frac{ 1 }{ 2 } } = 1 + a_{ s } C_{ F } \frac{ 1 }{ \epsilon } + \mathcal{ O } \! \left( \alpha_{ s }^{ 2 } \right) \! ,
\label{vwilson_selfenergy_nlo_explicit}
\end{equation}
Note that there are also IR divergences from the limit $ L \rightarrow \pm \infty $, which we collectively encode in a renormalization constant $ Z_{ W } $ (see also Eq. (\ref{renormalized_lp_baryon_qtmd_correlator_def})). These divergences come directly from the QTMD correlator, and cancel out in Eqs. (\ref{lp_baryon_qtmd_correlator_factorized}), (\ref{lp_baryon_qtmd_correlator_factorized_longitudinal_momentum_space}). Since $ \bar{ n }^{ 2 } = 0 $, the self-energy of the lightlike link is zero, and therefore the full renormalization constant is
\begin{align}
& Z_{ \Psi 1; \overline{ \text{MS} } } \! \left( \frac{ \sqrt{ - v^{ 2 } } \delta^{ - } }{ \mu v^{ - } } \right) = 1 - \frac{ a_{ s } C_{ F } }{ \epsilon^{ 2 } } \nonumber \\
& + \frac{ a_{ s } C_{ F } }{ \epsilon } \left( 1 + \ln \! \left( \frac{ \sqrt{ - v^{ 2 } } \delta^{ - } }{ \mu v^{ - } } \right)^{ \! 2 } \right) + \mathcal{ O } \! \left( a_{ s }^{ 2 } \right) \! .
\label{baryon_lfwf_lattice_uv_renormalization_constant_nlo_explicit}
\end{align}

\subsection{Factorization of the Soft Factor}
\label{factor_soft_factor_subsec}

At LO, the soft factor (\ref{baryon_lfwf_soft_factor}) reduces to the identity, and we conclude the section by demonstrating a completely factorized structure at next-to-leading order. The NLO corrections are computed in complete analogy with the lattice factor, yielding
\begin{align}
& S_{ \text{NLO} } \! \left( \left\{ b \right\} \! ; \delta^{ + }, \delta^{ - } \right) = - 2 a_{ s } \frac{ C_{ F } }{ N - 1 } \sum_{ \substack{ q, w = 1 \\ w \neq q } }^{ N } \left( \frac{ - \left( b_{ q } - b_{ w } \right)^{ 2 } }{ 4 } \right)^{ \epsilon } \nonumber \\
& \times \Gamma \! \left( - \epsilon \right) \left( - \psi \! \left( - \epsilon \right) - \gamma_{ E } + \ln \! \left( \frac{ - \left( b_{ q } - b_{ w } \right)^{ 2 } }{ 4 } 2 \delta^{ + } \delta^{ - } \right) \right) \! .
\label{nlo_baryon_lfwf_soft_factor}
\end{align}
Note that the rapidity-divergent term is linear in $ \ln \delta^{ + } \delta^{ - } $, therefore we can separate the contributions from $ n $-directed and $ \bar{ n } $-directed Wilson lines.

In the $ \overline{ \text{MS} } $ renormalization scheme and explicitly up to next-to-leading order, we have
\begin{widetext}
\begin{alignat}{2}
  S_{ \overline{ \text{MS} } } \! \left( \left\{ b \right\} \! ; \delta^{ 2 } \right)
& = 1
&
& - 2 a_{ s } \frac{ C_{ F } }{ N - 1 } \sum_{ \substack{ q, w = 1 \\ w \neq q } }^{ N }
	\left( \vphantom{ \ln \! \left( \frac{ - \left( b_{ q } - b_{ w } \right)^{ 2 } }{ 4 } \right) } \right.
\frac{ 1 }{ \epsilon^{ 2 } } - \left( \frac{ 1 }{ \epsilon } + \ln \! \left( \frac{ - \left( b_{ q } - b_{ w } \right)^{ 2 } }{ 4 } \mu^{ 2 } \right) \right) \ln \! \left( \frac{ 2 \delta^{ + } \delta^{ - } }{ \mu^{ 2 } } \right) \nonumber \\
&
&
& - \frac{ 1 }{ 2 } \ln^{ 2 } \! \left( \frac{ - \left( b_{ q } - b_{ w } \right)^{ 2 } }{ 4 } \mu^{ 2 } \right) - \frac{ \pi^{ 2 } }{ 12 }
	\left. \vphantom{ \ln \! \left( \frac{ - \left( b_{ q } - b_{ w } \right)^{ 2 } }{ 4 } \right) } \right)
+ \mathcal{ O } \! \left( a_{ s }^{ 2 } \right) \! .
\label{baryon_lfwf_soft_factor_nlo_explicit}
\end{alignat}
\end{widetext}
Introducing the momentum scales $ \nu_{ q }^{ \pm } $, we can isolate the rapidity-divergent contributions associated with $ n $-directed and $ \bar{ n } $-directed Wilson lines, i.e.,
\begin{equation}
R_{ \overline{ \text{MS} } } \! \left( \delta^{ \pm }, \left\{ \nu_{ q }^{ \pm } \right\} \right) = \prod_{ q = 1 }^{ N } R_{ \overline{ \text{MS} } } \! \left( \frac{ \delta^{ \pm } }{ \nu_{ q }^{ \pm } } \right) \! ,
\label{baryon_lfwf_soft_rapidity_divergent_correction_nlo_explicit}
\end{equation}
with
\begin{alignat}{1}
  R_{ \overline{ \text{MS} } } \! \left( \frac{ \delta^{ \pm } }{ \nu_{ q }^{ \pm } } \right) = 1
& + 2 a_{ s } \frac{ C_{ F } }{ N - 1 } \sum_{ \substack{ w = 1 \\ w \neq q } }^{ N } \ln \! \left( \frac{ - \left( b_{ q } - b_{ w } \right)^{ 2 } }{ 4 } \mu^{ 2 } \right)  \nonumber \\
& \times \ln \! \left( \frac{ \delta^{ \pm } }{ \nu_{ q }^{ \pm } } \right) + \mathcal{ O } \! \left( a_{ s }^{ 2 } \right) \! .
\label{baryon_lfwf_soft_leg_rapidity_divergent_correction_nlo_explicit}
\end{alignat}
Similarly, the renormalization constants associated with collinear divergences are
\begin{equation}
Z_{ R; \overline{ \text{MS} } } \! \left( \delta^{ \pm }, \left\{ \nu_{ q }^{ \pm } \right\} \right) = \prod_{ q = 1 }^{ N } Z_{ R; \overline{ \text{MS} } } \! \left( \frac{ \delta^{ \pm } }{ \nu_{ q }^{ \pm } } \right) \! .
\label{baryon_lfwf_soft_collinear_renormalization_constant_nlo_explicit}
\end{equation}
with
\begin{equation}
Z_{ R; \overline{ \text{MS} } } \! \left( \frac{ \delta^{ \pm } }{ \nu_{ q }^{ \pm } } \right) = 1 + 2 a_{ s } C_{ F } \frac{ 1 }{ \epsilon } \ln \! \left( \frac{ \delta^{ \pm } }{ \nu_{ q }^{ \pm } } \right) + \mathcal{ O } \! \left( a_{ s }^{ 2 } \right) \! .
\label{baryon_lfwf_soft_leg_collinear_renormalization_constant_nlo_explicit}
\end{equation}
The UV-divergent part is
\begin{equation}
Z_{ S; \overline{ \text{MS} } } \! \left( \left\{ \nu_{ q }^{ 2 } \right\} \right) = \prod_{ q = 1 }^{ N } Z_{ S; \overline{ \text{MS} } } \! \left( \nu_{ q }^{ 2 } \right) \! ,
\label{baryon_lfwf_soft_uv_renormalization_constant_nlo_explicit}
\end{equation}
with
\begin{equation}
Z_{ S; \overline{ \text{MS} } } \! \left( \nu_{ q }^{ 2 } \right) = 1 - 2 a_{ s } C_{ F } \left( \frac{ 1 }{ \epsilon^{ 2 } } - \frac{ 1 }{ \epsilon } \ln \! \left( \frac{ \nu_{ q }^{ 2 } }{ \mu^{ 2 } } \right) \right) + \mathcal{ O } \! \left( a_{ s }^{ 2 } \right) \! .
\label{baryon_lfwf_soft_leg_uv_renormalization_constant_nlo_explicit}
\end{equation}
The remaining finite part is
\begin{align}
& S_{ 0; \overline{ \text{MS} } } \! \left( \left\{ \nu_{ q }^{ 2 } \right\} \right) = 1 \nonumber \\
& + 2 a_{ s } \frac{ C_{ F } }{ N - 1 } \sum_{ \substack{ q, w = 1 \\ w \neq q } }^{ N } \ln \! \left( \frac{ - \left( b_{ q } - b_{ w } \right)^{ 2 } }{ 4 } \mu^{ 2 } \right) \ln \! \left( \frac{ \nu_{ q }^{ 2 } }{ \mu^{ 2 } } \right) \nonumber \\
& + \frac{ 1 }{ 2 } \ln^{ 2 } \! \left( \frac{ - \left( b_{ q } - b_{ w } \right)^{ 2 } }{ 4 } \mu^{ 2 } \right) + \frac{ \pi^{ 2 } }{ 12 } + \mathcal{ O } \! \left( a_{ s }^{ 2 } \right) \! .
\label{baryon_lfwf_soft_finite_part_nlo_explicit}
\end{align}
With the quantities introduced in Eqs. (\ref{baryon_lfwf_soft_rapidity_divergent_correction_nlo_explicit})--(\ref{baryon_lfwf_soft_finite_part_nlo_explicit}), we can express the soft factor in a completely factorized form, i.e.,
\begin{alignat}{2}
  S_{ \overline{ \text{MS} } } \! \left( \delta^{ 2 } \right)
& ={}
&
& \left( \prod_{ c = \pm } R_{ \overline{ \text{MS} } } \! \left( \delta^{ c }, \left\{ \nu_{ q }^{ c } \right\} \right) Z_{ R; \overline{ \text{MS} } } \! \left( \delta^{ c }, \left\{ \nu_{ q }^{ c } \right\} \right) \right) \nonumber \\
&
&
& \times Z_{ S; \overline{ \text{MS} } } \! \left( \left\{ \nu_{ q }^{ 2 } \right\} \right) S_{ 0; \overline{ \text{MS} } } \! \left( \left\{ \nu_{ q }^{ 2 } \right\} \right) \nonumber \\
& ={}
&
& R_{ \overline{ \text{MS} } } \! \left( \delta^{ + }, \left\{ \nu_{ q }^{ + } \right\} \right) R_{ \overline{ \text{MS} } } \! \left( \delta^{ - }, \left\{ \nu_{ q }^{ - } \right\} \right) \nonumber \\
&
&
& \times Z_{ S; \overline{ \text{MS} } }^{ N } \! \left( \delta^{ 2 } \right) S_{ 0; \overline{ \text{MS} } } \! \left( \left\{ \nu_{ q }^{ 2 } \right\} \right) \! .
\label{baryon_lfwf_soft_factor_factorized}
\end{alignat}

\section{Cancellation of Divergences and Physical Light-Front Wave Function}
\label{cancel_divs_n_physical_lfwf_sec}

We collect the results from the previous sections to verify, at leading power and next-to-leading order, the cancellation of divergences from the factorization of the QTMD correlator. The result will be a well-defined physical LFWF, independently of the residual lattice factor.

From now on, we will explicitly distinguish bare quantities from their renormalized counterparts. For the QTMD correlator, we have
\begin{equation}
\Omega_{ v; B; \text{bare} } \! \left( \left\{ x_{ q }, b_{ q } \right\} \right) = Z_{ W } Z_{ J }^{ 3 } \Omega_{ v; B } \! \left( \left\{ x_{ q }, b_{ q } \right\} \! , \mu, L \right) \! ,
\label{renormalized_lp_baryon_qtmd_correlator_def}
\end{equation}
where $ Z_{ W } $ renormalizes the $ v $-directed Wilson lines close to infinity, and each $ Z_{ J } $ renormalizes a remaining semicompact operator (\ref{quark_op_vwilsoned_to_infinity}). The same factor $ Z_{ W } $ appears in the renormalization of the lattice factor, and therefore cancels in Eqs. (\ref{lp_baryon_qtmd_correlator_factorized}), (\ref{lp_baryon_qtmd_correlator_factorized_longitudinal_momentum_space}) (see also discussion after Eq. (\ref{vwilson_selfenergy_nlo_explicit})). On the other hand, we have
\begin{align}
& \frac{ \Psi_{ \text{bare} } \! \left( \left\{ b_{ q } \right\} \right) C_{ 1 } \! \left( x_{ 1 } \right) C_{ 1 } \! \left( x_{ 2 } \right) C_{ 1 } \! \left( x_{ 3 } \right) \Phi_{ 1 1 1 ; \text{bare} } \! \left( \left\{ x_{ q }, b_{ q } \right\} \right) }{ S \! \left( \left\{ b_{ q } \right\} \right) } \nonumber \\
& = \frac{ 1 }{ Z_{ S } \! \left( \left\{ \nu_{ q }^{ 2 } \right\} \right) } \frac{ R \! \left( \delta^{ - }, \left\{ \nu_{ q }^{ - } \right\} \right) Z_{ \Psi 1 }^{ N } \! \left( \frac{ \sqrt{ - v^{ 2 } } \delta^{ - } }{ \mu v^{ - } } \right) }{ R \! \left( \delta^{ - }, \left\{ \nu_{ q }^{ - } \right\} \right) Z_{ R } \! \left( \delta^{ - }, \left\{ \nu_{ q }^{ - } \right\} \right) } \nonumber \\
& \times \frac{ R \! \left( \delta^{ + }, \left\{ \nu_{ q }^{ + } \right\} \right) \left( \prod_{q = 1}^{ N } Z_{ \Phi 1 } \! \left( \frac{ \delta^{ + } }{ p_{ q }^{ + } } \right) \right) }{ R \! \left( \delta^{ + }, \left\{ \nu_{ q }^{ + } \right\} \right) Z_{ R } \! \left( \delta^{ + }, \left\{ \nu_{ q }^{ + } \right\} \right) } \nonumber \\
& \times \frac{ \Psi \! \left( \left\{ b_{ q } \right\} \! , \mu, \left\{ \nu_{ q }^{ - } \right\} \right) C_{ 1 } C_{ 1 } C_{ 1 } \Phi_{ 1 1 1 } \! \left( \left\{ x_{ q }, b_{ q } \right\} \! , \mu, \left\{ \nu_{ q }^{ + } \right\} \right) }{ S_{ 0 } \! \left( \left\{ \nu_{ q }^{ 2 } \right\} \right) },
\label{baryon_qtmd_correlator_longitudinal_momentum_space_factorized_renormalized}
\end{align}
where, in the $ \overline{ \text{MS} } $ renormalization scheme, the divergences of the lattice factor are given by Eqs. (\ref{baryon_lfwf_lattice_rapidity_divergent_correction_nlo_explicit}), (\ref{baryon_lfwf_lattice_uv_renormalization_constant_nlo_explicit}), while those of the LFWF are given by Eqs. (\ref{lp_baryon_lfwf_rapidity_divergent_correction_nlo_explicit}), (\ref{lp_quark_uv_renormalization_constant_nlo_explicit}), and we used the factorized form (\ref{baryon_lfwf_soft_factor_factorized}) of the soft factor. Comparing the rapidity-divergent contributions in the numerator with those of the soft factor in Eq. (\ref{baryon_lfwf_soft_rapidity_divergent_correction_nlo_explicit}), we see that they cancel out exactly, thus justifying the use of the same notation. The collinear divergences in each $ Z_{ \Psi 1 }, Z_{ \Phi 1 } $ are instead canceled out by the corresponding renormalization factor (\ref{baryon_lfwf_soft_leg_collinear_renormalization_constant_nlo_explicit}), which trades the dependence on $ \delta^{ \pm } $ for $ \nu_{ q }^{ \pm } $. As a result, we are left with
\begin{align}
& \frac{ \prod_{ q = 1 }^{ N } Z_{ \Psi 1 } \! \left( \frac{ \sqrt{ - v^{ 2 } } \nu_{ q }^{ - } }{ \mu v^{ - } } \right) Z_{ \Phi 1 } \! \left( \frac{ \nu_{ q }^{ + } }{ p_{ q }^{ + } } \right) }{ Z_{ S } \! \left( \left\{ \nu_{ q }^{ 2 } \right\} \right)  } \frac{ \Psi C_{ 1 } C_{ 1 } C_{ 1 } \Phi_{ 1 1 1 } }{ S_{ 0 } \! \left( \left\{ \nu_{ q }^{ 2 } \right\} \right) } \nonumber \\
& = \left( \prod_{ q = 1 }^{ N } \frac{ Z_{ \Psi 1 } \! \left( \frac{ \sqrt{ - v^{ 2 } } \nu_{ q }^{ - } }{ \mu v^{ - } } \right) }{ Z_{ S }^{ \frac{ 1 }{ 2 } } \! \left( \nu_{ q }^{ 2 } \right) } \frac{ Z_{ \Phi 1 } \! \left( \frac{ \nu_{ q }^{ + } }{ p_{ q }^{ + } } \right) }{ Z_{ S }^{ \frac{ 1 }{ 2 } } \! \left( \nu_{ q }^{ 2 } \right) } \right) \nonumber \\
& \times \frac{ \Psi }{ S_{ 0 }^{ \frac{ 1 }{ 2 } } \! \left( \left\{ \nu_{ q }^{ 2 } \right\} \right) } C_{ 1 } C_{ 1 } C_{ 1 } \frac{ \Phi_{ 1 1 1 } }{ S_{ 0 }^{ \frac{ 1 }{ 2 } } \! \left( \left\{ \nu_{ q }^{ 2 } \right\} \right) } \nonumber \\
& = \left( \prod_{ q = 1 }^{ N } Z_{ \Psi 1; \text{sub} } Z_{ \Phi 1; \text{sub} } \right) \Psi_{ \text{sub} } C_{ 1 } C_{ 1 } C_{ 1 } \Phi_{ 1 1 1; \text{sub} } ,
\label{subtracted_quantities_intro}
\end{align}
where in the last step we have introduced the subtracted renormalization constants, as well as the subtracted LFWF and lattice factor. If each quark carries a fraction $ x_{ q } $ of the longitudinal momentum $ P^{ + } $ of the baryon, for each leg of the LFWF the subtracted renormalization constant is
\begin{align}
& Z_{ \Phi 1; \text{sub}; \overline{ \text{MS} } } \! \left( \frac{ \zeta_{ q } }{ \mu^{ 2 } } \right) = 1  + \frac{ a_{ s } C_{ F } }{ \epsilon^{ 2 } } \nonumber \\
& + \frac{ a_{ s } C_{ F } }{ \epsilon } \left( \frac{ 3 }{ 2 } - \ln \! \left( \frac{ \zeta_{ q } }{ \mu^{ 2 } } \right) \right) \nonumber \\
& - \frac{ 2 a_{ s } C_{ F } }{ \epsilon } \ln \! \left( i \text{sign} \! \left( L \right) \text{sign} \! \left( x_{ q } \right) \right) + \mathcal{ O } \! \left( a_{ s }^{ 2 } \right) \! ,
\label{lp_quark_uv_renormalization_constant_sub_nlo_explicit}
\end{align}
with
\begin{equation}
\zeta_{ q } = 2 \left( x_{ q } P^{ + } \right)^{ 2 } \frac{ \nu_{ q }^{ - } }{ \nu_{ q }^{ + } }.
\label{zetaq}
\end{equation}
For each leg of the lattice factor, the renormalization constant is instead
\begin{equation}
Z_{ \Psi 1; \text{sub}; \overline{ \text{MS} } } \! \left( \frac{ \bar{ \zeta }_{ q } }{ \mu^{ 2 } } \right) = 1 + a_{ s } C_{ F } \frac{ 1 }{ \epsilon } \left( 1 - \ln \! \left( \frac{ \bar{ \zeta }_{ q } }{ \mu^{ 2 } } \right) \right) + \mathcal{ O } \! \left( a_{ s }^{ 2 } \right) \! ,
\label{baryon_lfwf_lattice_uv_renormalization_constant_sub_nlo_explicit}
\end{equation}
with
\begin{equation}
\bar{ \zeta }_{ q } = 2 \frac{ \left( \mu v^{ - } \right)^{ 2 } }{ - v^{ 2 } } \frac{ \nu_{ q }^{ + } }{ \nu_{ q }^{ - } }.
\label{zetabarq}
\end{equation}
Note that the scales $ \zeta_{ q } $ and $ \bar{ \zeta }_{ q } $ are Lorentz invariant, and we have
\begin{alignat}{1}
  \Phi_{ 1 1 1; \text{sub} } \! \left( \left\{ x_{ q }, b_{ q } \right\} \right)
& = \Phi_{ 1 1 1; \text{sub} } \! \left( \left\{ x_{ q }, b_{ q } \right\} \! , \mu, \left\{ \zeta_{ q } \right\} \right) \nonumber \\
& = \frac{ \Phi_{ 1 1 1 } \! \left( \left\{ x_{ q }, b_{ q } \right\} \! , \mu, \left\{ \nu_{ q }^{ + } \right\} \right) }{ S_{ 0 }^{ \frac{ 1 }{ 2 } } \! \left( \left\{ \nu_{ q }^{ 2 } \right\} \right) } \! ,
\label{sub_lfwf_def} \\
  \Psi_{ \text{sub} } \! \left( \left\{ b_{ q } \right\} \right)
& = \Psi_{ \text{sub} } \! \left( \left\{ b_{ q } \right\} \! , \mu, \left\{ \bar{ \zeta }_{ q } \right\} \right) \nonumber \\
& = \frac{ \Psi \! \left( \left\{ b_{ q } \right\} \! , \mu, \left\{ \nu_{ q }^{ - } \right\} \right) }{ S_{ 0 }^{ \frac{ 1 }{ 2 } } \! \left( \left\{ \nu_{ q }^{ 2 } \right\} \right) } \! .
\label{sub_lattice_def}
\end{alignat}

To prove that each leg of the factorized QTMD correlator is individually finite, we need the divergent part of the NLO CF (\ref{lp_nlo_cf_longitudinal_momentum_space}), which, in the $ \overline{ \text{MS} } $ renormalization scheme, is
\begin{align}
& \text{pole} \! \left[ C_{ 1, \text{NLO} } \! \left( x \right) \right]_{ \overline{ \text{MS} } } \nonumber \\
& = \text{pole} \! \left[ 2 a_{ s } C_{ F } \frac{ 1 - \epsilon }{ 1 - 2 \epsilon } \Gamma \! \left( - \epsilon \right) \Gamma \! \left( 2 \epsilon \right) \left( \frac{ - v^{ 2 } \mu^{ 2 } e^{ - \gamma_{ E } } e^{ 2 \gamma_{ E } } }{ \left( i s s_{ x } 2 \lvert x \rvert v^{ - } P^{ + } \right)^{ 2 } } \right)^{ \! \epsilon } \right] \nonumber \\
& = - \frac{ a_{ s } C_{ F } }{ \epsilon^{ 2 } } - \frac{ a_{ s } C_{ F } }{ \epsilon } \left( 1 + \ln \! \left( \frac{ - v^{ 2 } \mu^{ 2 } }{ \left( 2 x v^{ - } P^{ + } \right)^{ 2 } } \right) \right) \nonumber \\
& + \frac{ 2 a_{ s } C_{ F } }{ \epsilon } \ln \! \left( i \text{sign} \! \left( L \right) \text{sign} \! \left( x \right) \right) \! .
\label{nlo_lp_cf_longitudinal_momentum_space_pole}
\end{align}
We also need the renormalization constant for the semicompact operator (\ref{quark_op_vwilsoned_to_infinity}), i.e.~\cite{Rodini:2022wic},
\begin{equation}
Z_{ J; \overline{ \text{MS} } } = 1 + \frac{ 3 }{ 2 } a_{ s } C_{ F } \frac{ 1 }{ \epsilon } + \mathcal{ O } \! \left( a_{ s }^{ 2 } \right) \! .
\label{quark_op_vwilsoned_to_infinity_renormalization_constant_nlo_explicit}
\end{equation}
Combining Eqs. (\ref{lp_quark_uv_renormalization_constant_sub_nlo_explicit}), (\ref{baryon_lfwf_lattice_uv_renormalization_constant_sub_nlo_explicit}), (\ref{nlo_lp_cf_longitudinal_momentum_space_pole}), (\ref{quark_op_vwilsoned_to_infinity_renormalization_constant_nlo_explicit}), and using
\begin{equation}
\zeta_{ q } \bar{ \zeta }_{ q } = \frac{ \left( 2 \mu v^{ - } x P^{ + } \right)^{ 2 } }{ - v^{ 2 } } = \frac{ \left( 2 x v^{ - } P^{ + } \right)^{ 2 } \mu^{ 4 } }{ - v^{ 2 } \mu^{ 2 } }
\label{zetaq_times_zetabarq}
\end{equation}
for every $ q = 1, 2, ..., N $, we prove the complete cancellation of divergences up to next-to-leading order, i.e.,
\begin{align}
& \left( Z_{ \Phi 1; \text{sub}; \overline{ \text{MS} } } \! \left( \frac{ \zeta_{ q } }{ \mu^{ 2 } } \right) Z_{ \Psi 1; \text{sub}; \overline{ \text{MS} } } \! \left( \frac{ \bar{ \zeta }_{ q } }{ \mu^{ 2 } } \right) Z_{ J; \overline{ \text{MS} } }^{ - 1 } \right)_{ \text{NLO} } \nonumber \\
& + \text{pole} \! \left[ C_{ 1, \text{NLO} } \! \left( x_{ q } \right) \right]_{ \overline{ \text{MS} } } = 0.
\label{cancellation_of_factorization_divergences}
\end{align}

After subtracting the poles, the remaining finite CF is given by
\begin{widetext}
\begin{alignat}{2}
  C_{ 1, \text{fin}; \overline{ \text{MS} } } \! \left( x \right)
& = 1
&
& - a_{ s } C_{ F } \left( 2 + \frac{ 5  \pi^{ 2 } }{ 12 } + \ln \! \left( \frac{ - v^{ 2 } \mu^{ 2 } }{ \left( i s s_{ x } 2 \lvert x \rvert v^{ - } P^{ + } \right)^{ 2 } } \right) + \frac{ 1 }{ 2 } \ln^{ 2 } \! \left( \frac{ - v^{ 2 } \mu^{ 2 } }{ \left( i s s_{ x } 2 \lvert x \rvert v^{ - } P^{ + } \right)^{ 2 } } \right) \right) + \mathcal{ O } \! \left( a_{ s }^{ 2 } \right) \nonumber \\
& = 1
&
& - a_{ s } C_{ F } \left(
	\vphantom{ \frac{ 1 }{ 2 } \ln^{ 2 } \! \left( \frac{ - v^{ 2 } \mu^{ 2 } }{ \left( 2 x v^{ - } P^{ + } \right)^{ 2 } } \right) } \right. 2 + \frac{ 5 \pi^{ 2 } }{ 12 } + \ln \! \left( \frac{ - v^{ 2 } \mu^{ 2 } }{ \left( 2 x v^{ - } P^{ + } \right)^{ 2 } }  \right) + \frac{ 1 }{ 2 } \ln^{ 2 } \! \left( \frac{ - v^{ 2 } \mu^{ 2 } }{ \left( 2 x v^{ - } P^{ + } \right)^{ 2 } } \right) \nonumber \\
&
&
& - 2 \ln \! \left( i s s_{ x } \right) + \frac{ 1 }{ 2 } \left( - 2 \ln \! \left( i s s_{ x } \right) \right)^{ 2 } \left. 
	\vphantom{ \frac{ 1 }{ 2 } \ln^{ 2 } \! \left( \frac{ - v^{ 2 } \mu^{ 2 } }{ \left( 2 x v^{ - } P^{ + } \right)^{ 2 } } \right) } \right) + \mathcal{ O } \! \left( a_{ s }^{ 2 } \right) \! .
\label{lp_cf_finite_nlo_explicit_complogs}
\end{alignat}
\end{widetext}
Since
\begin{equation}
2 \ln \! \left( i s s_{ x } \right) = 2 s s_{ x } i \frac{ \pi }{ 2 } = s s_{ x } i \pi,
\label{times2_log_i_s_sx}
\end{equation}
we can separate the real part of the CF from the imaginary part, i.e.,
\begin{align}
& C_{ 1, \text{fin}; \overline{ \text{MS} } } \! \left( x \right) = 1 + a_{ s } C_{ F } \left( \frac{ \pi^{ 2 } }{ 12 } - 2 \right) \nonumber \\
& - a_{ s } C_{ F } \left( \ln \! \left( \frac{ - v^{ 2 } \mu^{ 2 } }{ \left( 2 x v^{ - } P^{ + } \right)^{ 2 } }  \right) + \frac{ 1 }{ 2 } \ln^{ 2 } \! \left( \frac{ - v^{ 2 } \mu^{ 2 } }{ \left( 2 x v^{ - } P^{ + } \right)^{ 2 } } \right) \right) \nonumber \\
& + a_{ s } C_{ F } i \pi \text{sign} \! \left( L \right) \text{sign} \! \left( x \right) + \mathcal{ O } \! \left( a_{ s }^{ 2 } \right) \! .
\label{lp_cf_finite_nlo_explicit}
\end{align}
Therefore, the renormalized QTMD correlator factorizes as
\begin{alignat}{1}
  \Omega_{ v; B } \! \left( \left\{ x_{ q }, b_{ q } \right\} \! , \mu, L \right) ={}
& \Psi_{ \text{sub} } \! \left( \left\{ b_{ q } \right\} \! , \mu, \left\{ \bar{ \zeta }_{ q } \right\} \right) \mathcal{ C }_{ 1 1 1 } \! \left( \left\{ x_{ q } \right\} \right) \nonumber \\
& \times \Phi_{ 1 1 1; \text{sub} } \! \left( \left\{ x_{ q }, b_{ q } \right\} \! , \mu, \left\{ \zeta_{ q } \right\} \right) \! , 
\label{renormalized_lp_baryon_qtmd_correlator_factorized}
\end{alignat}
where, in the $ \overline{ \text{MS} } $ renormalization scheme and explicitly up to next-to-leading order, the CF for the baryon subtracted LP LFWF is
\begin{align}
& \mathcal{ C }_{ 1 1 1; \overline{ \text{MS} } } \! \left( \left\{ x_{ q } \right\} \right) = \prod_{ q = 1 }^{ 3 } C_{ 1, \text{fin}; \overline{ \text{MS} } } \! \left( x_{ q } \right) = 1 + a_{ s } C_{ F } \left( \frac{ \pi^{ 2 } }{ 4 } - 6 \right) \nonumber \\
& - a_{ s } C_{ F } \sum_{ q = 1 }^{ 3 } \ln \! \left( \frac{ - v^{ 2 } \mu^{ 2 } }{ \left( 2 x_{ q } v^{ - } P^{ + } \right)^{ 2 } } \right) \nonumber \\
& - a_{ s } C_{ F } \sum_{ q = 1 }^{ 3 } \frac{ 1 }{ 2 } \ln^{ 2 } \! \left( \frac{ - v^{ 2 } \mu^{ 2 } }{ \left( 2 x_{ q } v^{ - } P^{ + } \right)^{ 2 } } \right) \nonumber \\
& + a_{ s } C_{ F } i \pi \text{sign} \! \left( L \right) \sum_{ q = 1 }^{ 3 } \text{sign} \! \left( x_q \right) + \mathcal{ O } \! \left( a_{ s }^{ 2 } \right) \! .
\label{lp_baryon_lfwf_cf}
\end{align}
Our results naturally extend those in Ref.~\cite{Rodini:2022wic}, with a contribution from every possible pair of quarks. Since in our case there is no contraction between Dirac adjoints, the imaginary parts of the individual CFs do not cancel in the overall coefficient.

\section{Scale Evolution}
\label{scale_evo_sec}

The subtracted LFWF depends on three mass scales associated to rapidity divergences, i.e., $ \zeta_{ q } $ with $ q = 1, 2, 3 $, while the subtracted lattice factor depends on the $ \bar{ \zeta }_{ q } $, and both quantities also depend on the UV renormalization scale $ \mu $. We now study how the LFWF evolves with these scales, showing that the evolution equations with respect to each of them are independent. The solution takes an exponential form, with the rapidity-scale evolutions governed by generalizations of the Collins--Soper kernel.

From Eqs. (\ref{baryon_qtmd_correlator_longitudinal_momentum_space_factorized_renormalized}), (\ref{subtracted_quantities_intro}), we have
\begin{alignat}{1}
  \frac{ \Phi_{ 1 1 1 ; \text{bare} } \Psi_{ \text{bare} } }{ S } ={}
& \left( \prod_{ q' = 1 }^{ 3 } Z_{ \Phi 1; \text{sub} } \! \left( \frac{ \zeta_{ q' } }{ \mu^{ 2 } } \right) Z_{ \Psi 1; \text{sub} } \! \left( \frac{ \bar{ \zeta }_{ q' } }{ \mu^{ 2 } } \right) \right) \nonumber \\
& \times \Phi_{ 1 1 1; \text{sub} } \! \left( \mu, \left\{ \zeta_{ q } \right\} \right) \Psi_{ \text{sub} } \! \left( \mu, \left\{ \bar{ \zeta }_{ q } \right\} \right) \! .
\label{lp_baryon_qtmd_correlator_factorized_bare_vs_sub}
\end{alignat}

We take the logarithmic derivative  with respect to $ \mu^{ 2 } $ of the logarithm of Eq. (\ref{lp_baryon_qtmd_correlator_factorized_bare_vs_sub}), at fixed value of the rapidity scales. The left-hand side does not depend on the renormalization scales, and, since only the LFWF depends on $ \left\{ \zeta_{ q } \right\} $ and only the lattice factor depends on $ \left\{ \bar{ \zeta }_{ q } \right\} $, we find two independent evolution equations. For the LFWF, we have
\begin{equation}
\mu^{ 2 } \frac{ \partial }{ \partial \mu^{ 2 } } \Phi_{ 1 1 1; \text{sub} } \! \left( \mu, \left\{ \zeta_{ q } \right\} \right) = \sum_{ q' = 1 }^{ 3 } \gamma_{ \Phi 1 } \! \left( x_{ q' }, \frac{ \zeta_{ q' } }{ \mu^{ 2 } } \right) \Phi_{ 1 1 1; \text{sub} } ,
\label{lp_baryon_lfwf_evolution_wrt_mu}
\end{equation}
where, from Eq. (\ref{lp_quark_uv_renormalization_constant_sub_nlo_explicit}), the anomalous dimensions is
\begin{align}
& \gamma_{ \Phi 1 } \! \left( x_{ q }, \frac{ \zeta_{ q } }{ \mu^{ 2 } } \right) = - \mu^{ 2 } \frac{ 1 }{ Z_{ \Phi 1; \text{sub} } } \frac{ \partial }{ \partial \mu^{ 2 } } Z_{ \Phi 1; \text{sub} } \nonumber \\
& \overset{ \overline{ \text{MS} } }{=} a_{ s } C_{ F } \left( \frac{ 3 }{ 2 } - \ln \frac{ \zeta_{ q } }{ \mu^{ 2 } } \right) \nonumber \\
& - 2 a_{ s } C_{ F } \ln \! \left( i \text{sign} \! \left( L \right) \text{sign} \! \left( x_{ q } \right) \right) + \mathcal{ O } \! \left( a_{ s }^{ 2 } \right) \! .
\label{quark_anomalous_dimension_nlo_explicit}
\end{align}
Similarly, for the lattice factor we find
\begin{equation}
\mu^{ 2 } \frac{ \partial }{ \partial \mu^{ 2 } } \Psi_{ \text{sub} } \! \left( \mu, \left\{ \bar{ \zeta }_{ q } \right\} \right) = \sum_{ q' = 1 }^{ 3 } \gamma_{ \Psi 1 } \! \left( \frac{ \bar{ \zeta }_{ q' } }{ \mu^{ 2 } } \right) \Psi_{ \text{sub} } \! \left( \mu, \left\{ \bar{ \zeta }_{ q } \right\} \right) \! ,
\label{baryon_lfwf_lattice_evolution_wrt_mu}
\end{equation}
where, using Eq. (\ref{baryon_lfwf_lattice_uv_renormalization_constant_sub_nlo_explicit}), the anomalous dimension is
\begin{align}
& \gamma_{ \Psi 1 } \! \left( \frac{ \bar{ \zeta }_{ q } }{ \mu^{ 2 } } \right) = - \mu^{ 2 } \frac{ 1 }{ Z_{ \Psi 1; \text{sub} } } \frac{ \partial }{ \partial \mu^{ 2 } } Z_{ \Psi 1; \text{sub} } \nonumber \\
& \overset{ \overline{ \text{MS} } }{=} + a_{ s } C_{ F } \left( 1 - \ln \frac{ \bar{ \zeta }_{ q } }{ \mu^{ 2 } } \right) + \mathcal{ O } \! \left( a_{ s }^{ 2 } \right) \! .
\label{baryon_lfwf_lattice_anomalous_dimension}
\end{align}

From Eq. (\ref{zetaq_times_zetabarq}), we have that $ \zeta_{ q }, \bar{ \zeta }_{ q } $ are not independent when $ \mu $ is fixed. Therefore, to derive the rapidity-scale evolution of the LFWF,  we consider
\begin{equation}
\Phi_{ 1 1 1 ; \text{bare} } = \left( \prod_{ q' = 1 }^{ 3 } R \! \left( \frac{ \delta^{ + } }{ \nu_{ q' }^{ + } } \right) Z_{ \Phi 1 } \! \left( \frac{ \delta^{ + } }{ p_{ q' }^{ + } } \right) \right) \Phi_{ 1 1 1 } \! \left( \mu, \left\{ \nu_{ q }^{ + } \right\} \right) \! .
\label{lp_baryon_lfwf_factorized_renormalized}
\end{equation}
From Eq. (\ref{zetaq}), we have
\begin{equation}
\zeta_{ q } \frac{ \partial }{ \partial \zeta_{ q } } = - \nu_{ q }^{ + } \frac{ \partial }{ \partial \nu_{ q }^{ + } }.
\label{derivative_wrt_zetaq_vs_nuplus}
\end{equation}
Taking the logarithmic derivative with respect to $ \nu^{ + } $ of Eq. (\ref{lp_baryon_lfwf_factorized_renormalized}), we have
\begin{align}
& \nu_{ q }^{ + } \frac{ \partial \ln \! \left( R \! \left( \frac{ \delta^{ + } }{ \nu_{ q }^{ + } } \right) \right) }{ \partial \nu_{ q }^{ + } } \Phi_{ 1 1 1 } \! \left( \mu, \left\{ \nu_{ q }^{ + } \right\} \right)\nonumber \\
& = - \nu_{ q }^{ + } \frac{ \partial }{ \partial \nu_{ q }^{ + } } \Phi_{ 1 1 1 } \! \left( \mu, \left\{ \nu_{ q }^{ + } \right\} \right) \nonumber \\
& = - S_{ 0 }^{ \frac{ 1 }{ 2 } } \! \left( \left\{ \nu_{ q }^{ 2 } \right\} \right) \nu_{ q }^{ + } \frac{ \partial }{ \partial \nu_{ q }^{ + } } \Phi_{ 1 1 1; \text{sub} } \! \left( \mu, \left\{ \zeta_{ q } \right\} \right) \nonumber \\
& - \nu_{ q }^{ + } \frac{ \partial S_{ 0 }^{ \frac{ 1 }{ 2 } } \! \left( \left\{ \nu_{ q }^{ 2 } \right\} \right) }{ \partial \nu_{ q }^{ + } } \Phi_{ 1 1 1; \text{sub} } \! \left( \mu, \left\{ \zeta_{ q } \right\} \right) \! ,
\label{lp_baryon_lfwf_renormalized_evolution_wrt_zetaq}
\end{align}
where in the last equality we used Eq. (\ref{sub_lfwf_def}). Deriving the second equality in Eq. (\ref{baryon_lfwf_soft_factor_factorized}) and using Eqs. (\ref{baryon_lfwf_soft_rapidity_divergent_correction_nlo_explicit}), (\ref{baryon_lfwf_soft_leg_rapidity_divergent_correction_nlo_explicit}), we also have
\begin{align}
& \nu_{ q }^{ \pm } \frac{ \partial S_{ 0 }^{ \frac{ 1 }{ 2 } } \! \left( \left\{ \nu_{ q }^{ 2 } \right\} \right) }{ \partial \nu_{ q }^{ \pm } } = \frac{ 1 }{ S_{ 0 }^{ \frac{ 1 }{ 2 } } \! \left( \left\{ \nu_{ q }^{ 2 } \right\} \right) } \frac{ \nu_{ q }^{ \pm } }{ 2 } \frac{ \partial S_{ 0 } \! \left( \left\{ \nu_{ q }^{ 2 } \right\} \right) }{ \partial \nu_{ q }^{ \pm } } \nonumber \\
& = S_{ 0 }^{ \frac{ 1 }{ 2 } } \! \left( \left\{ \nu_{ q }^{ 2 } \right\} \right) \frac{ \nu_{ q }^{ \pm } }{ 2 } \frac{ \partial \ln \! \left( S_{ 0 } \! \left( \left\{ \nu_{ q }^{ 2 } \right\} \right) \right) }{ \partial \nu_{ q }^{ \pm } } \nonumber \\
& = - S_{ 0 }^{ \frac{ 1 }{ 2 } } \! \left( \left\{ \nu_{ q }^{ 2 } \right\} \right) \frac{ \nu_{ q }^{ \pm } }{ 2 } \frac{ \partial \ln \! \left( R \! \left( \frac{ \delta^{ \pm } }{ \nu_{ q }^{ \pm } } \right) \right) }{ \partial \nu_{ q }^{ \pm } } .
\label{derivative_of_soft_finite_vs_rapidity_renormalization}
\end{align}
Combining Eqs. (\ref{derivative_wrt_zetaq_vs_nuplus})--(\ref{derivative_of_soft_finite_vs_rapidity_renormalization}), we find
\begin{equation}
\zeta_{ q } \frac{ \partial }{ \partial \zeta_{ q } } \Phi_{ 1 1 1; \text{sub} } \! \left( \mu, \left\{ \zeta_{ q } \right\} \right) = \frac{ \nu_{ q }^{ + } }{ 2 } \frac{ \partial \ln \! \left( R \! \left( \frac{ \delta^{ + } }{ \nu_{ q }^{ + } } \right) \right) }{ \partial \nu_{ q }^{ + } } \Phi_{ 1 1 1; \text{sub} } .
\label{lp_baryon_lfwf_evolution_wrt_zetaq}
\end{equation}
Note that, using Eq. (\ref{baryon_lfwf_soft_leg_rapidity_divergent_correction_nlo_explicit}), we have
\begin{equation}
- \frac{ \nu_{ q }^{ + } }{ 2 } \frac{ \partial \ln \! \left( R \! \left( \frac{ \delta^{ + } }{ \nu_{ q }^{ + } } \right) \right) }{ \partial \nu_{ q }^{ + } } = \frac{ 1 }{ 4 } \sum_{ \substack{ w = 1 \\ w \neq q } }^{ 3 } \mathcal{ D } \! \left( \left( b_{ q } - b_{ w } \right)^{ 2 } \! , \mu \right) \! ,
\label{lp_baryon_lfwf_rapidity_renormalization_vs_cs_kernel}
\end{equation}
where
\begin{equation}
\mathcal{ D }_{ \overline{ \text{MS} } } \! \left( b^{ 2 }, \mu \right) = + 2 a_{ s } C_{ F } \ln \! \left( \frac{ -  b^{ 2 } }{ 4 } \mu^{ 2 } \right) + \mathcal{ O } \! \left( a_{ s }^{ 2 } \right)
\label{cs_kernel_nlo_explicit}
\end{equation}
is the one-loop perturbative part of the standard Collins--Soper kernel.

The integrability conditions follow from the commutativity of the partial derivatives. From Eq. (\ref{lp_baryon_lfwf_evolution_wrt_zetaq}) and the fact that the left-hand side of Eq. (\ref{lp_baryon_lfwf_rapidity_renormalization_vs_cs_kernel}) is independent of the rapidity scales, second order derivatives with respect to  these scales vanish, and therefore the corresponding first-order derivatives commute. Moreover, consistency between Eqs. (\ref{lp_baryon_lfwf_evolution_wrt_mu}) and (\ref{lp_baryon_lfwf_evolution_wrt_zetaq}) requires
\begin{equation}
\zeta_{ q } \frac{ \partial \gamma_{ \Phi 1 } \! \left( x_{ q }, \frac{ \zeta_{ q } }{ \mu^{ 2 } } \right) }{ \partial \zeta_{ q } } = - \mu^{ 2 } \frac{ \partial }{ \partial \mu^{ 2 } } \left( - \frac{ \nu_{ q }^{ + } }{ 2 } \frac{ \partial \ln \! \left( R \! \left( \frac{ \delta^{ + } }{ \nu_{ q }^{ + } } \right) \right) }{ \partial \nu_{ q }^{ + } } \right) \! ,
\label{lp_baryon_lfwf_scale_evolution_integrability_condition}
\end{equation}
which is indeed satisfied, since, using Eqs. (\ref{quark_anomalous_dimension_nlo_explicit}), (\ref{lp_baryon_lfwf_rapidity_renormalization_vs_cs_kernel}), (\ref{cs_kernel_nlo_explicit}), we find
\begin{align}
& - \mu^{ 2 } \frac{ \partial }{ \partial \mu^{ 2 } } \left( - \frac{ \nu_{ q }^{ + } }{ 2 } \frac{ \partial \ln \! \left( R \! \left( \frac{ \delta^{ + } }{ \nu_{ q }^{ + } } \right) \right) }{ \partial \nu_{ q }^{ + } } \right) \nonumber \\
& = - \frac{ 1 }{ 4 } \sum_{ \substack{ w = 1 \\ w \neq q } }^{ 3 } \mu^{ 2 } \frac{ \partial \mathcal{ D } \! \left( \left( b_{ q } - b_{ w } \right)^{ 2 }, \mu \right) }{ \partial \mu^{ 2 } } \nonumber \\
  ={}
& - \frac{ 1 }{ 4 } \sum_{ \substack{ w = 1 \\ w \neq q } }^{ 3 } \frac{ \Gamma_{ \text{cusp} } }{ 2 } = - \frac{ 1 }{ 4 } \Gamma_{ \text{cusp} } \nonumber \\
& \overset{ \overline{ \text{MS} } }{=} - a_{ s } C_{ F } + \mathcal{ O } \! \left( a_{ s }^{ 2 } \right) \overset{ \overline{ \text{MS} } }{=} \zeta_{ q } \frac{ \partial \gamma_{ \Phi 1 } \! \left( x_{ q }, \frac{ \zeta_{ q } }{ \mu^{ 2 } } \right) }{ \partial \zeta_{ q } },
\label{lp_baryon_lfwf_scale_evolution_integrability_condition_nlo_explicit}
\end{align}
where $ \Gamma_{ \text{cusp} } $ is the standard cusp anomalous dimension, and in the explicit expression we only kept nonzero terms in the limit $ \epsilon \rightarrow 0 $.

Finally, combining Eqs. (\ref{lp_baryon_lfwf_evolution_wrt_mu}), (\ref{lp_baryon_lfwf_evolution_wrt_zetaq}), (\ref{lp_baryon_lfwf_rapidity_renormalization_vs_cs_kernel}), we find
\begin{alignat}{1}
  d \ln \Phi_{ 1 1 1; \text{sub} } \! \left( \mu, \left\{ \zeta_{ q } \right\} \right) ={}
& \sum_{ q = 1 }^{ 3 } \gamma_{ \Phi 1 } \! \left( x_{ q }, \frac{ \zeta_{ q } }{ \mu^{ 2 } } \right) \frac{ d \mu^{ 2 } }{ \mu^{ 2 } } \nonumber \\
& - \frac{ 1 }{ 4 } \sum_{ \substack{ q, w = 1 \\ w \neq q } }^{ 3 } \mathcal{ D } \! \left( \left( b_{ q } - b_{ w } \right)^{ 2 } \! , \mu \right) \frac{ d \zeta_{ q } }{ \zeta_{ q } },
\label{lp_baryon_lfwf_total_differential}
\end{alignat}
whose solution can be written as
\begin{widetext}
\begin{equation}
\Phi_{ 1 1 1; \text{sub} } \! \left( \mu, \left\{ \zeta_{ q } \right\} \right) = \Phi_{ 1 1 1; \text{sub} } \! \left( \mu_{ 0 }, \left\{ \zeta_{ q, 0 } \right\} \right) P \exp \int \! \left( \sum_{ q = 1 }^{ 3 } \gamma_{ \Phi 1 } \! \left( x_{ q }, \frac{ \zeta_{ q } }{ \mu^{ 2 } } \right) \frac{ d \mu^{ 2 } }{ \mu^{ 2 } } - \frac{ 1 }{ 4 } \sum_{ \substack{ q, w = 1 \\ w \neq q } }^{ 3 } \mathcal{ D } \! \left( \left( b_{ q } - b_{ w } \right)^{ 2 } \! , \mu \right) \frac{ d \zeta_{ q } }{ \zeta_{ q } } \right) \! ,
\label{lp_baryon_lfwf_scale_evolution}
\end{equation}
\end{widetext}
where the path-ordered exponential is evaluated between some initial scales $ \left( \mu_{ 0 }, \left\{ \zeta_{ q, 0 } \right\} \right) $ and scales $ \left( \mu, \left\{ \zeta_{ q } \right\} \right) $.

We are free to choose the path of integration in Eq. (\ref{lp_baryon_lfwf_scale_evolution}), e.g., we can evolve each $ \zeta_{ q } $ independently from all other scales, and in the end evolve $ \mu $ at fixed values of the rapidity scales. First, we have
\begin{align}
& \exp \! \int_{ \zeta_{ q, 0 } }^{ \zeta_{ q } } \! \frac{ d \zeta_{ q }' }{ \zeta_{ q }' } \left( - \frac{ 1 }{ 4 } \sum_{ \substack{ w = 1 \\ w \neq q } }^{ 3 } \mathcal{ D } \! \left( \left( b_{ q } - b_{ w } \right)^{ 2 } \! , \mu \right) \right) \nonumber \\
& = \left( \frac{ \zeta_{ q } }{ \zeta_{ q, 0 } } \right)^{ \! - \frac{ 1 }{ 4 } \sum_{ \substack{ w = 1 \\ w \neq q } }^{ 3 } \mathcal{ D } \left( \left( b_{ q } - b_{ w } \right)^{ 2 } \! , \mu \right) }.
\label{zetaq_evolution}
\end{align}
Therefore, we have
\begin{align}
& \frac{ \Phi_{ 1 1 1; \text{sub} } \! \left( \mu, \left\{ \zeta_{ q } \right\} \right) }{ \Phi_{ 1 1 1; \text{sub} } \! \left( \mu_{ 0 }, \left\{ \zeta_{ q, 0 } \right\} \right) } \nonumber \\
& = \prod_{ q = 1 }^{ 3 } \left( \frac{ \zeta_{ q } }{ \zeta_{ q, 0 } } \right)^{ \! - \frac{ 1 }{ 4 } \sum_{ \substack{ w = 1 \\ w \neq q } }^{ 3 } \mathcal{ D } \left( \left( b_{ q } - b_{ w } \right)^{ 2 } \! , \mu \right) } \exp \int \! \frac{ d \mu^{ 2 } }{ \mu^{ 2 } } \gamma_{ \Phi 1, q } .
\label{lp_baryon_lfwf_scale_evolution_zetaq_first}
\end{align}

\section{Conclusions}
\label{conclusions_sec}

In this work, we have developed a framework to connect light-front wave functions (LFWFs) of baryons such as the proton to equal-time correlators suitable for Lattice QCD simulations. Building on established factorization approaches for transverse-momentum-dependent (TMD) distributions and meson LFWFs, we have constructed the quasi-transverse-momentum-dependent (QTMD) correlator that can be matched to a light-front amplitude through a TMD operator expansion based on the background field method.

Focusing on the leading three-quark color-singlet LFWF, we have established its factorization theorem from the QTMD correlator up to next-to-leading order. At the bare level, the factorized expression consists of the LFWF and a residual lattice-dependent factor. Both develop additional ultraviolet (UV) and rapidity divergences compared to the original QTMD correlator, the latter being generated by interactions at infinity with lightlike Wilson lines. We have shown how these divergences are canceled one by one by an appropriate soft factor, allowing for a well-defined physical LFWF independently of the residual lattice factor.

The resulting LFWF is multiplicatively renormalizable quark leg by quark leg, and depends on a UV renormalization scale as well as one rapidity scale for each quark. The corresponding scale evolutions are independent, and can be organized within renormalization-group equations. In particular, the rapidity evolutions are governed by generalized Collins--Soper kernels. The evolution exponentiates and reflects the pairwise color correlations characteristic of baryons.

Although explicit calculations were carried out in a specific gauge, the resulting factorization and evolution structure is general, and naturally extends known results for TMD parton distributions and fragmentation functions to the richer color structure of a baryon.

The approach presented in this paper opens up several directions for future work. On the theoretical side, a detailed study of higher Fock components and their mixing under renormalization would help clarify how multi-parton correlations emerge in the proton wave function. It will also be important to investigate the stability of the factorization framework beyond next-to-leading order, and to explore possible simplifications for practical lattice implementations. On the numerical side, the next step is the explicit computation of the proposed correlators in lattice simulations. This requires a careful analysis of systematic effects such as finite-volume corrections, discretization errors, and the treatment of large hadron momenta. A quantitative extraction of proton LFWFs would allow direct comparisons with phenomenological models, and provide new input for the interpretation of experimental observables. More broadly, establishing a reliable method to determine baryon LFWFs from lattice QCD would strengthen the link between first-principle calculations and the parton description of hadron structure. Since many measurable quantities can be understood in terms of LFWFs, access to the wave functions themselves would provide a unified and more detailed picture of the internal dynamics of the proton.

\begin{appendix}

\section{Notations and Conventions}
\label{notations_n_conventions_sec}

In this appendix, we compile the definitions and conventions used throughout the manuscript, together with some useful relations.
 
We decompose a generic four-vector $ a = \left( a^{ + }, a^{ - }, \vec{ a }_{ \perp } \right) $ in terms of two lightlike vectors $ \bar{ n }, n $, with $ \bar{ n }^{ 2 } = 0 = n^{ 2 } $, $ \bar{ n } n = 1 $, and a transverse part $ a_{ \perp } = \left( 0, 0, \vec{ a }_{ \perp } \right) $, so that
\begin{equation}
a_{ \perp }^{ 2 } \leq 0, \quad \bar{ n } a_{ \perp } = 0 = n a_{ \perp }, \quad \bar{ n } a = a^{ - }, \quad n a = a^{ + }.
\label{sudakov_decomposition}
\end{equation}
The following operators acting on spinor fields form a complete set of orthonormal projectors:
\begin{align}
& 	\Lambda_{ + } = \frac{ 1 }{ 2 } \gamma^{ - } \gamma^{ + } = \frac{ 1 }{ \sqrt{ 2 } } \gamma^{ 0 } \gamma^{ + }, 
\label{lc_good_projector} \\	
& 	\Lambda_{ - } = \frac{ 1 }{ 2 } \gamma^{ + } \gamma^{ - } = \frac{ 1 }{ \sqrt{ 2 } } \gamma^{ 0 } \gamma^{ - },
\label{lc_bad_projector}
\end{align}
where $ \gamma^{ \pm }, \gamma^{ 0 } $ are Dirac matrices.

The propagator for a Dirac fermion in position space and $ D = 4 - 2 \epsilon $ dimensions is
\begin{equation}
\wick{ \c{ \psi }_{ i } \! \left( a \right) \c{ \overline{ \psi } }_{ j } \! \left( b \right) } = \frac{ i }{ 2 \pi^{ \frac{ D }{ 2 } } } \Gamma \! \left( 2 - \epsilon \right) \frac{ \slashed{ a } - \slashed{ b } }{ \left( - \left( a - b \right)^{ \! 2 } + i 0 \right)^{ \! 2 - \epsilon } } \delta_{ i j },
\label{dirac_propagator_position_space}
\end{equation}
while the gauge-boson propagator is
\begin{equation}
\wick{ \c{ A }_{ \mu }^{ I } \! \left( a \right) \c{ A }_{ \nu }^{ J } \! \left( b \right) } = \frac{ - 1 }{ 4 \pi^{ \frac{ D }{ 2 } } } \Gamma \! \left( 1 - \epsilon \right) \frac{ g_{ \mu \nu } }{ \left( - \left( a - b \right)^{ \! 2 } + i 0 \right)^{ \! 1 - \epsilon } } \delta^{ I J },
\label{gauge_propagator_position_space}
\end{equation}
where $ i, j, I, J $ are color indices, and $ \Gamma $ is the Euler Gamma function.

For a $ \text{SU} \! \left( N \right) $ gauge group with generators $ \left\{ t^{ A } \right\}_{ A = 1, 2, ..., N^{ 2 } - 1 } $ in the fundamental representation, the color algebra satisfies,
\begin{equation}
\sum_{ A = 1 }^{ N^{ 2 } - 1 } t^{ A }_{ i j } t^{ A }_{ k l} = \frac{ 1 }{ 2 } \! \left( \delta_{ i l } \delta_{ j k } - \frac{ 1 }{ N } \delta_{ i j } \delta_{ k l } \right) \! .
\label{product_of_generators_of_sun}
\end{equation}
We further define the quadratic Casimir multiplicative operator in the fundamental representation as
\begin{equation}
C_{ F } = \frac{ N^{ 2 } - 1 }{ 2 N }.
\label{cf}
\end{equation}

In dimensional regularization, with $ D = 4 - 2 \epsilon $ dimensions, we define
\begin{equation}
a_{ s } = \frac{ \alpha_{ s } }{ \left( 4 \pi \right)^{ 1 - \epsilon } } = \frac{ g^{ 2 } }{ \left( 4 \pi \right)^{ 2 - \epsilon } },
\label{as}
\end{equation}
where $ g $ is the strong coupling constant. In the $ \overline{ \text{MS} } $ renormalization scheme, the factor $ \left( 4 \pi \right)^{ \epsilon } $ cancels, and we extract a factor of $ \mu^{ \epsilon } $ from the coupling constant, with $ \mu $ a mass scale, but we keep the same notation for simplicity. We define the anomalous dimension associated with a renormalization constant $ Z $ as
\begin{equation}
\gamma = - \mu^{ 2 } \frac{ \partial \ln Z }{ \partial \mu^{ 2 } } \approx - \mu^{ 2 } \frac{ \partial Z }{ \partial \mu^{ 2 } } ,
\label{anomalous_dim_def}
\end{equation}
where the approximate equality follows from the perturbative expansion
\begin{equation}
Z = 1 + \mathcal{ O } \! \left( a_{ s } \right) \! .
\label{renormalization_constant_form}
\end{equation}

\section{Improved Contractions}

\begin{equation}
i g \int_{ 0 }^{ L } \! d \sigma  v^{ \mu }
\tikz[baseline = (Bmu.base), remember picture]
{
	\node[inner sep = 0] (Bmu) {$ \vphantom{ \slashed{ B } } B $};
}\vphantom{ B }_{ \mu } \! \left( \sigma v + b \right)
\tikz[baseline = (psi.base), remember picture]
{
	\node[inner sep = 0] (psi) {$ \vphantom{ \overline{ \psi } } \psi $};
} \! \left( y \right) i g \int \! d^{ D } z
\tikz[baseline = (psibar.base), remember picture]
{
	\node[inner sep = 0] (psibar) {$ \overline{ \psi } $};
}
\tikz[overlay, remember picture]
{
	\path (psi.north)    -- ++(0, 0.03) coordinate (psinorth);
	\path (psi.north)    -- ++(0, 0.19) coordinate (topofpsi);
	\path (psibar.north) -- ++(0, 0.03) coordinate (psibarnorth);
	\path (psibar.north) -- ++(0, 0.19) coordinate (topofpsibar);
	\draw (psinorth) -- (topofpsi) -- (topofpsibar) -- (psibarnorth);
}
\tikz[baseline = (Bnu.base), remember picture]
{
	\node[inner sep = 0] (Bnu) {$ \slashed{ B } $};
}
\tikz[overlay, remember picture]
{
	\path (Bmu.north) -- ++(0, 0.03) coordinate (Bmunorth);
	\path (Bmu.north) -- ++(0, 0.31) coordinate (topofBmu);
	\path (Bnu.north) -- ++(0, 0.03) coordinate (Bnunorth);
	\path (Bnu.north) -- ++(0, 0.31) coordinate (topofBnu);
	\draw (Bmunorth) -- (topofBmu) -- (topofBnu) -- (Bnunorth);
} q_{ \bar{ n } } \! \left( z \right) \! .
\label{lp_nlo_effective_quark_current_contraction_v2}
\end{equation}

\begin{align}
& \varepsilon_{ i j k } \delta_{ i i' } \delta_{ k k' } \xi_{ \bar{ n }, j' } \! \left( y_{ 2 } \right) \xi_{ \bar{ n }, k' } \nonumber \\
& \times i g \int_{ y_{ 2 }^{ - } }^{ L } \! d \sigma n^{ \mu }
\tikz[baseline = (Bmu.base), remember picture]
{
	\node[inner sep = 0] (Bmu) {$ \vphantom{ \slashed{ B } } B $};
}\vphantom{ B }_{ \mu, j j' } \! \left( \sigma n + b_{ 2 } \right) \tikz[baseline = (psi.base), remember picture]
{
	\node[inner sep = 0] (psi) {$ \vphantom{ \overline{ \psi } } \psi $};
}_{ i' } \! \left( y_{ 1 } \right) i g \int \! d^{ D } z
\tikz[baseline = (psibar.base), remember picture]
{
	\node[inner sep = 0] (psibar) {$ \overline{ \psi } $};
}
\tikz[overlay, remember picture]
{
	\path (psi.north)    -- ++(0, 0.03) coordinate (psinorth);
	\path (psi.north)    -- ++(0, 0.19) coordinate (topofpsi);
	\path (psibar.north) -- ++(0, 0.03) coordinate (psibarnorth);
	\path (psibar.north) -- ++(0, 0.19) coordinate (topofpsibar);
	\draw (psinorth) -- (topofpsi) -- (topofpsibar) -- (psibarnorth);
}
\tikz[baseline = (Bnu.base), remember picture]
{
	\node[inner sep = 0] (Bnu) {$ \slashed{ B } $};
}
\tikz[overlay, remember picture]
{
	\path (Bmu.north) -- ++(0, 0.03) coordinate (Bmunorth);
	\path (Bmu.north) -- ++(0, 0.31) coordinate (topofBmu);
	\path (Bnu.north) -- ++(0, 0.03) coordinate (Bnunorth);
	\path (Bnu.north) -- ++(0, 0.31) coordinate (topofBnu);
	\draw (Bmunorth) -- (topofBmu) -- (topofBnu) -- (Bnunorth);
} q_{\bar{ n }  } \! \left( z \right) \! ,
\label{nlo_lp_baryon_lfwf_12_contraction_v2}
\end{align}

\begin{align}
& \varepsilon_{ i j k } \delta_{ i i' } \delta_{ k k'' } \delta_{ j' j'' } \varepsilon_{ i'' j'' k'' } v^{ \mu } \nonumber \\
& \times i g \int_{ 0 }^{ L } \! d \sigma
\tikz[baseline = (Bmu.base), remember picture]
{
	\node[inner sep = 0] (Bmu) {$ \vphantom{ \slashed{ B } } B $};
}\vphantom{ B }_{ \mu, i' i'' } \! \left( \sigma v + b_{ 1 } \right) i g \int_{ L }^{ 0 } \! d \chi \bar{ n }^{ \nu } \tikz[baseline = (Bnu.base), remember picture]
{
	\node[inner sep = 0] (Bnu) {$ \slashed{ B } $};
}\vphantom{ B }_{ \nu, j j' } \! \left( \chi \bar{ n } + b_{ 2 } \right) \! ,
\tikz[overlay, remember picture]
{
	\path (Bmu.north) -- ++(0, 0.03) coordinate (Bmunorth);
	\path (Bmu.north) -- ++(0, 0.31) coordinate (topofBmu);
	\path (Bnu.north) -- ++(0, 0.03) coordinate (Bnunorth);
	\path (Bnu.north) -- ++(0, 0.31) coordinate (topofBnu);
	\draw (Bmunorth) -- (topofBmu) -- (topofBnu) -- (Bnunorth);
}
\label{nlo_baryon_lfwf_lattice_12_contraction_v2}
\end{align}

\begin{equation}
\tikz[baseline = (psi.base), remember picture]
{
	\node[inner sep = 0] (psi) {$ \vphantom{ \overline{ \psi } } \psi $};
}_{ i } \! \left( a \right)
\tikz[baseline = (psibar.base), remember picture]
{
	\node[inner sep = 0] (psibar) {$ \overline{ \psi } $};
}_{ j } \! \left( b \right)
\tikz[overlay, remember picture]
{
	\path (psi.north)    -- ++(0, 0.03) coordinate (psinorth);
	\path (psi.north)    -- ++(0, 0.13) coordinate (topofpsi);
	\path (psibar.north) -- ++(0, 0.03) coordinate (psibarnorth);
	\path (psibar.north) -- ++(0, 0.13) coordinate (topofpsibar);
	\draw (psinorth) -- (topofpsi) -- (topofpsibar) -- (psibarnorth);
} = \frac{ i }{ 2 \pi^{ \frac{ D }{ 2 } } } \Gamma \! \left( 2 - \epsilon \right) \frac{ \slashed{ a } - \slashed{ b } }{ \left( - \left( a - b \right)^{ \! 2 } + i 0 \right)^{ \! 2 - \epsilon } } \delta_{ i j },
\label{dirac_propagator_position_space_v2}
\end{equation}

\begin{equation}
\tikz[baseline = (Amu.base), remember picture]
{
	\node[inner sep = 0] (Amu) {$ \vphantom{ A^{ I } } A $};
}\vphantom{ A }_{ \mu }^{ I } \! \left( a \right)
\tikz[baseline = (Anu.base), remember picture]
{
	\node[inner sep = 0] (Anu) {$ \vphantom{ A^{ J } } A $};
}\vphantom{ A }_{ \nu }^{ J } \! \left( b \right)
\tikz[overlay, remember picture]
{
	\path (Amu.north) -- ++(0, 0.03) coordinate (Amunorth);
	\path (Amu.north) -- ++(0, 0.13) coordinate (topofAmu);
	\path (Anu.north) -- ++(0, 0.03) coordinate (Anunorth);
	\path (Anu.north) -- ++(0, 0.13) coordinate (topofAnu);
	\draw (Amunorth) -- (topofAmu) -- (topofAnu) -- (Anunorth);
} = \frac{ - 1 }{ 4 \pi^{ \frac{ D }{ 2 } } } \Gamma \! \left( 1 - \epsilon \right) \frac{ g_{ \mu \nu } }{ \left( - \left( a - b \right)^{ \! 2 } + i 0 \right)^{ \! 1 - \epsilon } } \delta^{ I J },
\label{gauge_propagator_position_space_v2}
\end{equation}

\end{appendix}

\bibliography{connecting_baryon_lfwf_to_lattice_biblio}

\end{document}